\documentclass{aa}  

\usepackage{graphicx}
\usepackage{txfonts}
\usepackage{natbib}
\usepackage{multirow}
\usepackage{array}
\bibpunct{(}{)}{;}{a}{}{,}

\newcommand{\jdu}{\mbox{$J$=2$-$1}}

\newcommand{\juc}{\mbox{$J$=1$-$0}}

\newcommand{\kms}{\mbox{km\,s$^{-1}$}}

\newcommand{\lsim}{\raisebox{-.45ex}{$\stackrel{\sf <}{\scriptstyle\sf \sim}$}}
\newcommand{\gsim}{\raisebox{-.45ex}{$\stackrel{\sf >}{\scriptstyle\sf \sim}$}}
\newcommand{\aprop}{\raisebox{-.45ex}{$\stackrel{\propto}{\scriptstyle\sf \sim}$}}

\newcommand{\secp}{\mbox{\rlap{.}$''$}}
\begin{document}

   \title{$^{28}$SiO $v$=0 \juc\ emission from evolved stars} 


   \author{P. de Vicente\inst{1}
          \and
          V. Bujarrabal\inst{2}
	  \and
          A. D\'\i az-Pulido\inst{1}
          \and
          C. Albo\inst{1}
	  \and
          J. Alcolea\inst{3}
	  \and
          A. Barcia\inst{1}
          \and
          L. Barbas\inst{1}
          \and
          R. Bola\~no\inst{1}
          \and
          F. Colomer\inst{4}
          \and
          M.C. Diez\inst{1}
          \and
          J.D. Gallego\inst{1}
          \and
          J. G\'omez-Gonz\'alez\inst{4}
          \and
          I. L\'opez-Fern\'andez\inst{1}
          \and
          J.A. L\'opez-Fern\'andez\inst{1}
          \and
          J.A. L\'opez-P\'erez\inst{1}
          \and
          I. Malo\inst{1}
          \and
          A. Moreno\inst{1}
          \and
          M. Patino\inst{1}
          \and
          J.M. Serna\inst{1}
          \and
          F. Tercero\inst{1}
          \and
          B. Vaquero\inst{1}
}

   \institute{Observatorio de Yebes (IGN), Apartado 148, 19180, Yebes,  Spain \\     
\email{p.devicente@oan.es} 
        \and
Observatorio Astron\'omico Nacional (OAN-IGN), Apartado 112, E-28803 Alcal\'a de Henares, Spain
        \and
Observatorio Astron\'omico Nacional (OAN-IGN), Alfonso XII 3, E-28014, Madrid, Spain
        \and
Instituto Geogr\'afico Nacional, General Iba\~nez de Ibero 3, E-28003, Madrid, Spain
  }

   \date{Received August 11, 2015; accepted February 24, 2016}

 
  \abstract
   {}
{Observations of $^{28}$SiO $v$=0 \juc\ line emission (7-mm wavelength)
  from Asymptotic Giant Branch (AGB) stars show  in some cases peculiar profiles, composed of a central 
intense component plus a wider plateau. Very similar profiles have been
observed in CO lines from some AGB stars and most post-AGB nebulae and, in these cases, they are
clearly associated with the presence of conspicuous axial symmetry and bipolar dynamics. 
We aim to systematically study the profile shape of 
$^{28}$SiO $v$=0 \juc\ lines in evolved stars and to discuss the origin of the composite 
profile structure. 
}
   {We present observations of $^{28}$SiO $v$=0
     \juc\ emission in 28 evolved stars,  
including O-rich, C-rich, and S-type Mira-type variables, OH/IR stars,
semiregular long-period variables,  
red supergiants and one yellow hypergiant. Most objects were observed  in 
several epochs, over a total period of time of one and a half years. The observations were  
performed with the 40~m radio telescope of the Instituto Geogr\'afico Nacional (IGN) in Yebes, Spain.}
 {We find that the composite core plus plateau profiles are systematically present in O-rich Miras,  
OH/IR stars, and red supergiants. They are also found in one S-type
Mira ($\chi$ Cyg) and in two semiregular variables (X Her and RS Cnc) that are known to
show axial symmetry. In the other objects, the profiles are simpler and
similar to those observed  in other molecular lines. The composite structure
appears in the objects  in which SiO emission is thought to come from the very inner
circumstellar layers, prior to  
dust formation. The central spectral feature is found to be 
systematically composed of a number of narrow spikes, except for X Her
and RS Cnc, in which  it shows a smooth shape that is very similar to that observed in CO
emission. These spikes show a  significant (and mostly chaotic) time variation, while in all cases the
smooth components remain constant within the uncertainties. 
The profile shape could come from the superposition of standard wide profiles and a group of
weak maser spikes confined to the central spectral regions because of tangential amplification. 
Alternatively, we speculate that the very similar profiles detected in objects that are known to be conspicuously
axisymmetric, such as X Her and RS Cnc, and in O-rich Mira-type stars, such as IK {\rm T}au and TX Cam,
may be indicative of the systematic presence of a significant axial symmetry in the very inner 
circumstellar shells around AGB stars; such symmetry would be independent of the presence of weak maser effects 
in the central spikes.
}
   {}


   \maketitle
%

\section{Introduction}

SiO rotational line emission is often detected in evolved stars. These
lines are particularly intense in O-rich Asymptotic Giant Branch (AGB) circumstellar envelopes
(CSEs), but they are also detected in C-rich objects and other evolved stars. 
The low-$J$ vibrationally
excited ($v$$\geq$1) lines present intense maser emission in AGB stars,
particularly in O-rich Mira-type variables
\citep[][etc.]{buhl74,bujetal87,alc90,jewell91,deguchi10,cho12}. Their
profiles present some narrow spikes, distributed over a spectral
band of a few \kms. The maser emitting region is very small, consisting of 
a number of spots in a ring-like structure with a radius of
few stellar radii that are centered on the star \citep[e.g.][]{diamond94,
  gonidakis13, desmurs14}, this structure is due to the dominant 
tangential amplification. The variability of these masers is very
strong and, at least in regular pulsators, tightly follows the stellar
IR variability cycle (similar to the optical variability cycle, but with a phase shift of about
0.1--0.2); see e.g.\ \cite{pardo04}.

The $v$=0 \jdu\ and higher-J lines have also been observed in AGB and post-AGB
CSEs. It seems to be a typical thermal emission, which is much weaker than the
$v$$>$0 masers and shows wider profiles that are more or less comparable to
those observed in lines of other standard molecules, such as CO, HCN, etc.
\citep[e.g.][]{gonzalez03, bujetal89, shoier11}. In
O-rich AGB stars, the $v$=0 \jdu\ line probably originates in very
inner shells, within about 10$^{15}$ cm, as shown by interferometric
maps \citep{lucas92}, see also \cite{gonzalez03}. In this region, grains are still being formed and
the velocity gradient is still noticeable, which explains the presence
of silicon in the gas, not yet fully locked into grains, as well as the
observed triangular line profiles \citep[e.g.][]{bujetal89}. On the contrary in
C-rich AGB stars this line seems to come from more extended
shells and the profiles are very similar to those of the standard
molecules arising from the bulk of the envelope.

The $v$=0 rotational lines of the rare isotopes, $^{29}$SiO and
$^{30}$SiO, sometimes present spiky profiles that are attributed to
maser emission. These lines are weaker than $^{28}$SiO $v$$>$0 masers, but
are stronger than the $^{28}$SiO $v$=0 lines \citep{alc92}.

The $^{28}$SiO $v$=0 \juc\ emission from evolved stars, at 7-mm wavelength, has
been scarcely studied so far, since it is relatively weaker than the
$v$$>$0 masers. In some objects, it was found to show very peculiar
composite profiles \citep{jewell91} with a wide and smooth plateau and
a central spiky component that was proposed to be 
due to weak maser emission \citep{bobcla04}.  
Remarkably, similar composite profiles have been observed in 
  standard thermal lines, particularly in CO, 
coming from a rare class of objects, most of which are semiregular
variables; see \cite{knapp98}, \cite{libert10} and 
\cite{ccarrizo10}. In contrast to the case of the majority of AGB shells, in these objects 
the whole circumstellar envelope is 
strongly axisymmetric; it is an hourglass-like structure that consists of a
slowly expanding equatorial disk (at \lsim\ 5 
\kms) that is responsible for the central spectral feature, plus a somewhat
faster bipolar outflow (expanding at $\sim$ 10 
\kms).  The axial structure and line profiles found in these 
semiregular stars is very similar to what is typical of young planetary and 
protoplanetary nebulae \citep[ PNe, PPNe; see CO emission data in 
e.g.][]{bujetal01, alc07} although in post-AGB nebulae 
the involved velocities are significantly higher.  
The mechanism responsible for the spectacular evolution from AGB CSEs
to form PPNe and PNe is still under debate. It is thought to involve
the effects of a companion on the inner circumstellar regions
\citep[see][and references therein]{blackman14}. In this scenario, the
companion should also affect (at least) the inner shells in the AGB
phase, yielding some kind of axisymmetric structure. In some objects,
binarity is known to yield a large-scale axial symmetry in AGB CSEs
\citep[e.g.][]{ramstedt14}, which seems to be the case for the above mentioned
semiregular variables but certainly not for most AGB stars. 

In this paper, we present a wide survey of $^{28}$SiO
$v$=0 \juc\ emission from evolved stars of different kinds which also
includes observations in different epochs of a number of sources; we
show that such composite profiles systematically appear in O-rich and
S-type sources. We will suggest that the $^{28}$SiO $v$=0 line in
O-rich AGB stars is formed, as the other SiO lines, in very inner
circumstellar shells, and that the observed composite profiles may 
reflect the systematically axisymmetric structure of these regions, although at
large scales such effects would in some way disappear or be more
difficult to detect.

\begin{table}
\caption{Observed sources: used coordinates, LSR velocity and spectral type}.
Sources are ordered alphabetically within the types.
\scriptsize
\begin{tabular}{|l|ll|c|l|}
\hline 
 & & & & \\ 
name & R.A. (J2000) & dec.\ (J2000) & $V_ {\rm LSR}$ (\kms) & Sp.
type \\
\hline    
\multicolumn{5}{|c|}{} \\ 
\multicolumn{5}{|c|}{O-rich Mira-type variable stars} \\
\hline
 & & & & \\
RR Aql & 19:57:36.06 & $-$01:53:11.3 & $+$28.8  & M7.5e \\
TX Cam & 05:00:50.39 & $+$56:10:52.6 & $+$10.5 & M8.5e \\ 
R Cas & 23:58:24.87 & $+$51:23:19.7 & $+$25.5 & M6.5-9e \\ 
$o$ Cet & 02:19:20.79 & $-$02:58:39.5 & $+$46.5 & M1-9e \\ 
R Hya & 13:29:42.78 & $-$23:16:52.8 & $-$10 & M6-9e \\ 
R Leo & 09:47:33.49 & $+$11:25:43.7 & $-$1 & M7-9e \\ 
R LMi & 09:45:34.28 &  $+$34:30:42.8 & $+$0.5 & M6.5-9e \\ 
GX Mon & 06:52:47.04 & $+$08:25:19.2 & $-$10.5 & M9 \\ 
IK Tau & 03:53:28.87 & $+$11:24:21.7 & $+$34 & M9 \\ 
\hline  
\multicolumn{5}{|c|}{} \\ 
\multicolumn{5}{|c|}{S-type Mira-type variable stars} \\ 
\hline
 & & & & \\ 
W Aql & 19:15:23.347 & $-$07:02:50.3 & $-$27.5 & S6/6e \\ 
$\chi$ Cyg & 19:50:33.92 & $+$32:54:50.6 & $+$10 & S6-9/1-2e \\ 
\hline
\multicolumn{5}{|c|}{} \\ 
\multicolumn{5}{|c|}{OH/IR stars} \\ 
\hline
 & & & & \\ 
IRC\,$+$10011 & 01:06:25.98 & $+$12:35:53.0 & $+$9.5 & M8 \\  
OH\,26.5$+$0.6 & 18:37:32.51 & $-$05:23:59.2  & $+$29  & M  \\ 
OH\,44.8$-$2.3 & 19:21:36.52 & $+$09:27:56.5  & $-$71  & M  \\ 
\hline
\multicolumn{5}{|c|}{} \\ 
\multicolumn{5}{|c|}{C-rich Mira-type variable stars} \\ 
\hline
 & & & & \\ 
LP And & 23:34:27.53 & $+$43:33:01.2 & $-$19.5 & C8,3.5e \\  
CIT 6 & 10:16:02.27 & $+$30:34:18.6 & $+$4 & C4,3e  \\ 
IRC\,$+$10216 & 09:47:57.41 & $+$13:16:43.6 & $-$26 & C9,5e \\
\hline
\multicolumn{5}{|c|}{} \\ 
\multicolumn{5}{|c|}{O-rich semiregular pulsating stars} \\ 
\hline
 & & & & \\  
RX Boo & 14:24:11.63 & $+$25:42:13.4 & $+$1 & M7.5 \\ 
RS Cnc & 09:10:38.80 & $+$30:57:47.3 & $+$7.5 & M6S \\  
R Crt  & 11:00:33.85 & $-$18:19:29.6 & $+$11.5 & M7 \\ 
X Her  & 16:02:39.17 & $+$47:14:25.3 & $-$73 & M8 \\ 
W Hya  & 13:49:02.00 & $-$28:22:03.5 & $+$41 & M7.5-9e \\ 
RT Vir & 13:02:37.98 & $+$05:11:08.4 & $+$18 & M8 \\ 
\hline
\multicolumn{5}{|c|}{} \\ 
\multicolumn{5}{|c|}{O-rich red supergiant stars} \\ 
\hline
 & & & & \\  
VY CMa & 07:22:58.33 & $-$25:46:03.2 & $+$17 & M2.5-5Iae \\ 
NML Cyg & 20:46:25.54 & $+$40:06:59.4 & $-$1 & M7-8 \\ 
S Per & 02:22:51.71 & $+$58:35:11.4 & $-$41.5 & M3Iae \\ 
VX Sgr & 18:08:04.05 & $-$22:13:26.6 & $+$5 & M5/M6 \\ 
\hline
\multicolumn{5}{|c|}{} \\ 
\multicolumn{5}{|c|}{O-rich yellow hypergiant star} \\
\hline
 & & & & \\   
IRC\,$+$10420 & 19:26:48.095 & $+$11:21:16.74 & $+$76.5 & F8Ia+e \\ 
\hline
\end{tabular} 
\end{table}


   \begin{figure}
   \centering{\resizebox{8cm}{!}{
   \includegraphics{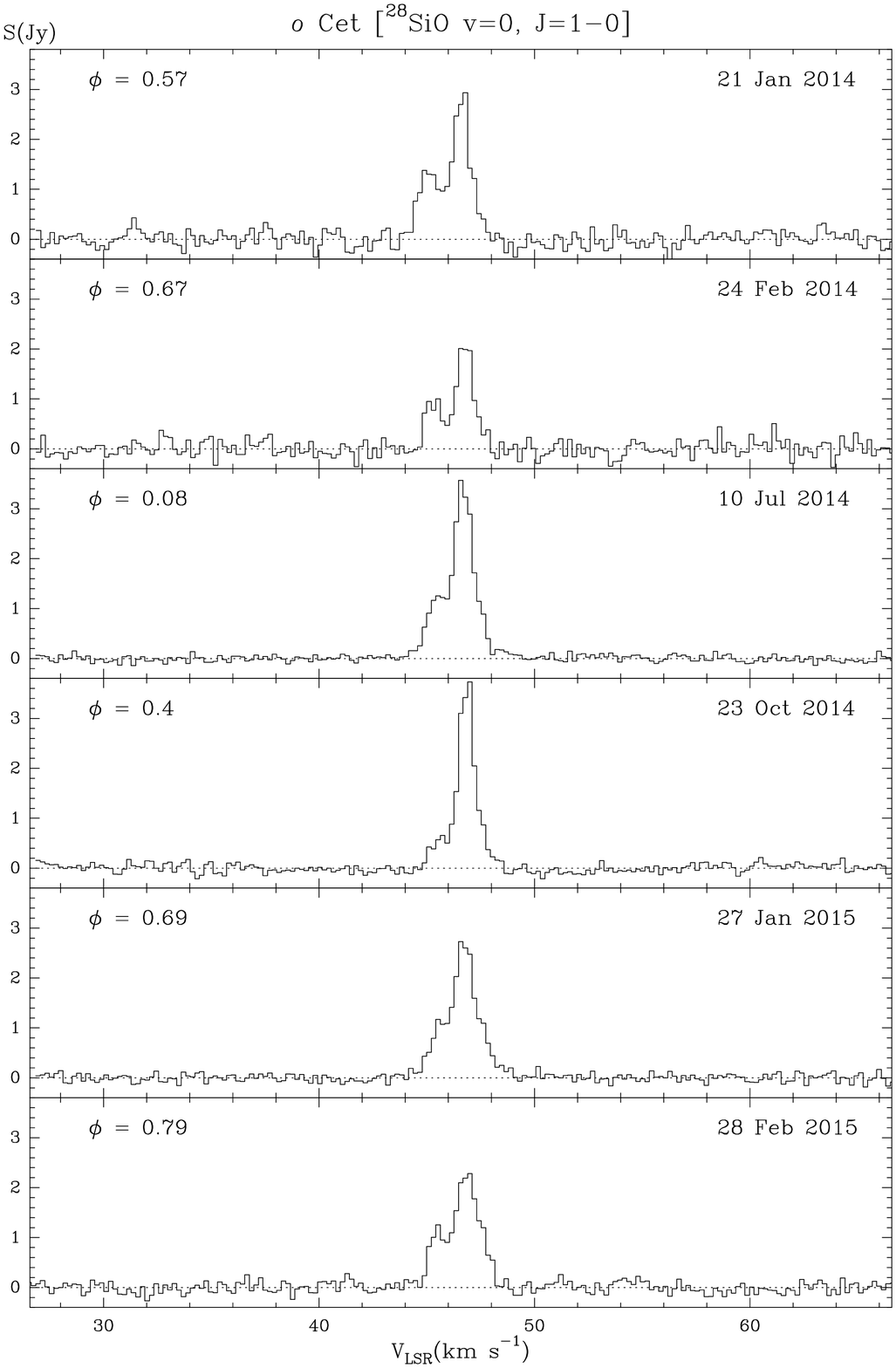}}}
   \caption{SiO $v$=0 \juc\ spectra in the O-rich Mira-type variable star
     $o$ Cet (Mira). The units and dates of the observations are
     indicated. The phase in the visible is shown in the upper left corner.}
    \label{f.omicet}%
    \end{figure}

   \begin{figure}
   \centering{\resizebox{8cm}{!}{
   \includegraphics{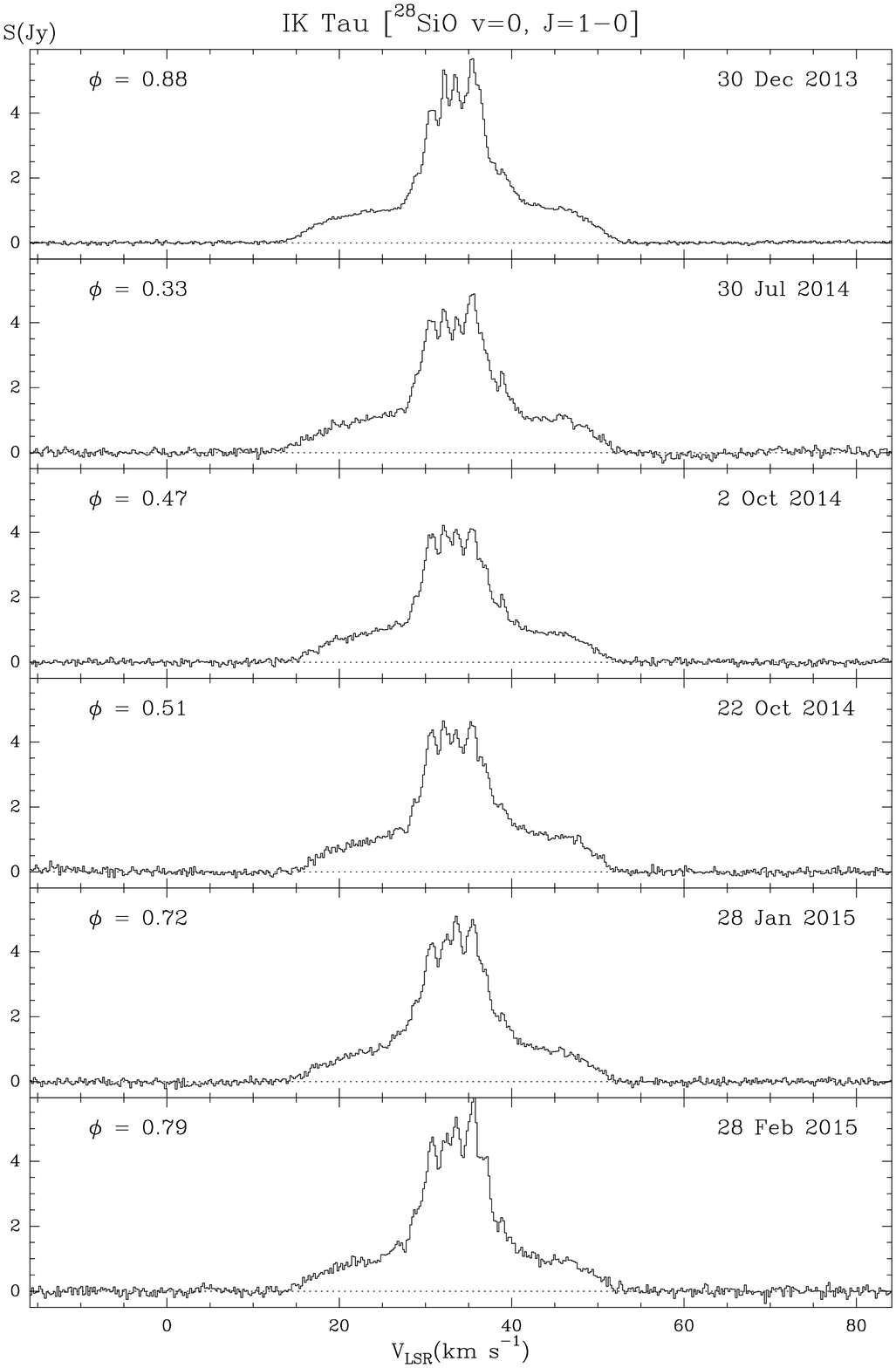}}}
   \caption{SiO $v$=0 \juc\ spectra in the O-rich Mira-type variable
     star IK Tau. The units and dates of the observations are
     indicated. The phase in the visible is shown in the upper left corner.}
    \label{f.iktau}%
    \end{figure}

   \begin{figure}
   \centering{\resizebox{8cm}{!}{
   \includegraphics{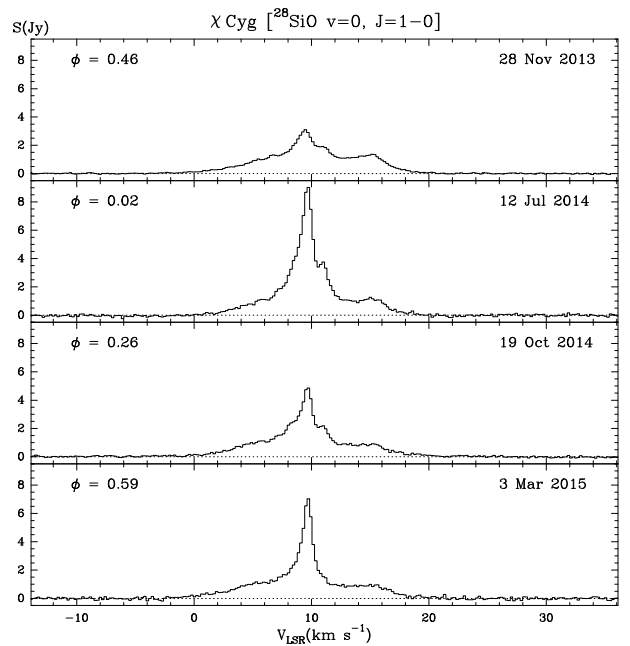}}}
   \caption{SiO $v$=0 \juc\ spectra in S-type Mira-type variable star
     $\chi$ Cyg. The units and dates of the observations are
     indicated. The phase in the visible is shown in the upper left corner.}
    \label{f.chicyg}%
    \end{figure}

   \begin{figure}
   \centering{\resizebox{8cm}{!}{
   \includegraphics{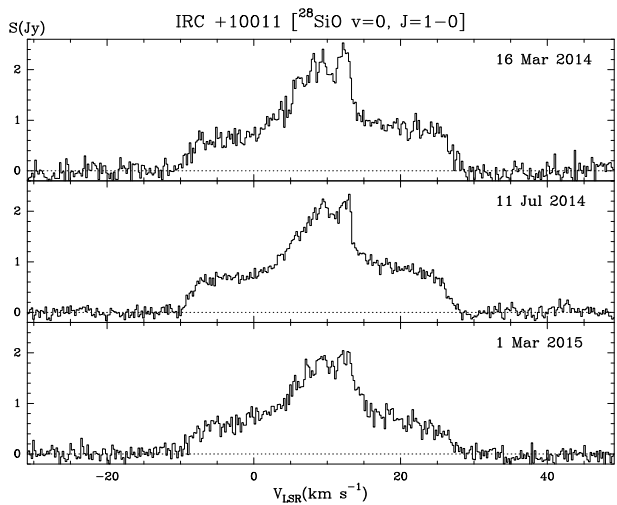}}}
   \caption{SiO $v$=0 \juc\ spectra in the OH/IR star IRC\,+10011. The
     units and dates of the observations are indicated.}
    \label{f.irc10011}%
    \end{figure}
   \begin{figure}
   \centering{\resizebox{8cm}{!}{
   \includegraphics{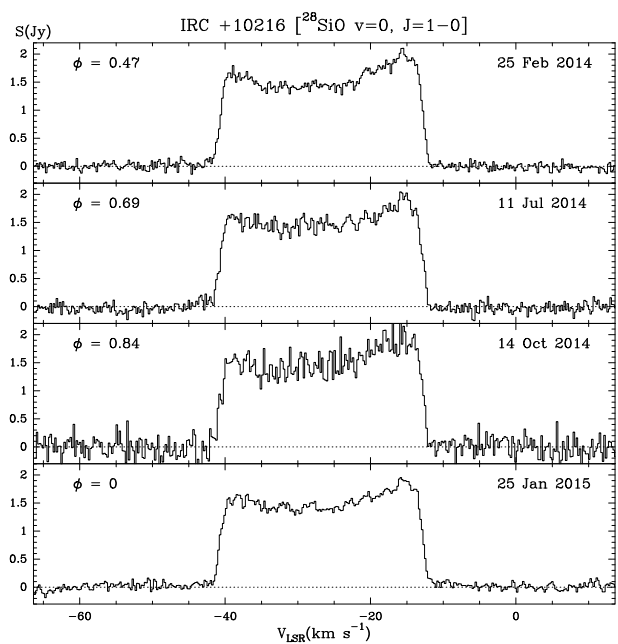}}}
   \caption{SiO $v$=0 \juc\ spectra in C-rich Mira-type variable star
     IRC\,+10216 (CW Leo). The units and dates of the observations are
     indicated. The phase in the visible is shown in the upper left corner.}
    \label{f.irc10216}%
    \end{figure}

   \begin{figure}
   \centering{\resizebox{8cm}{!}{
   \includegraphics{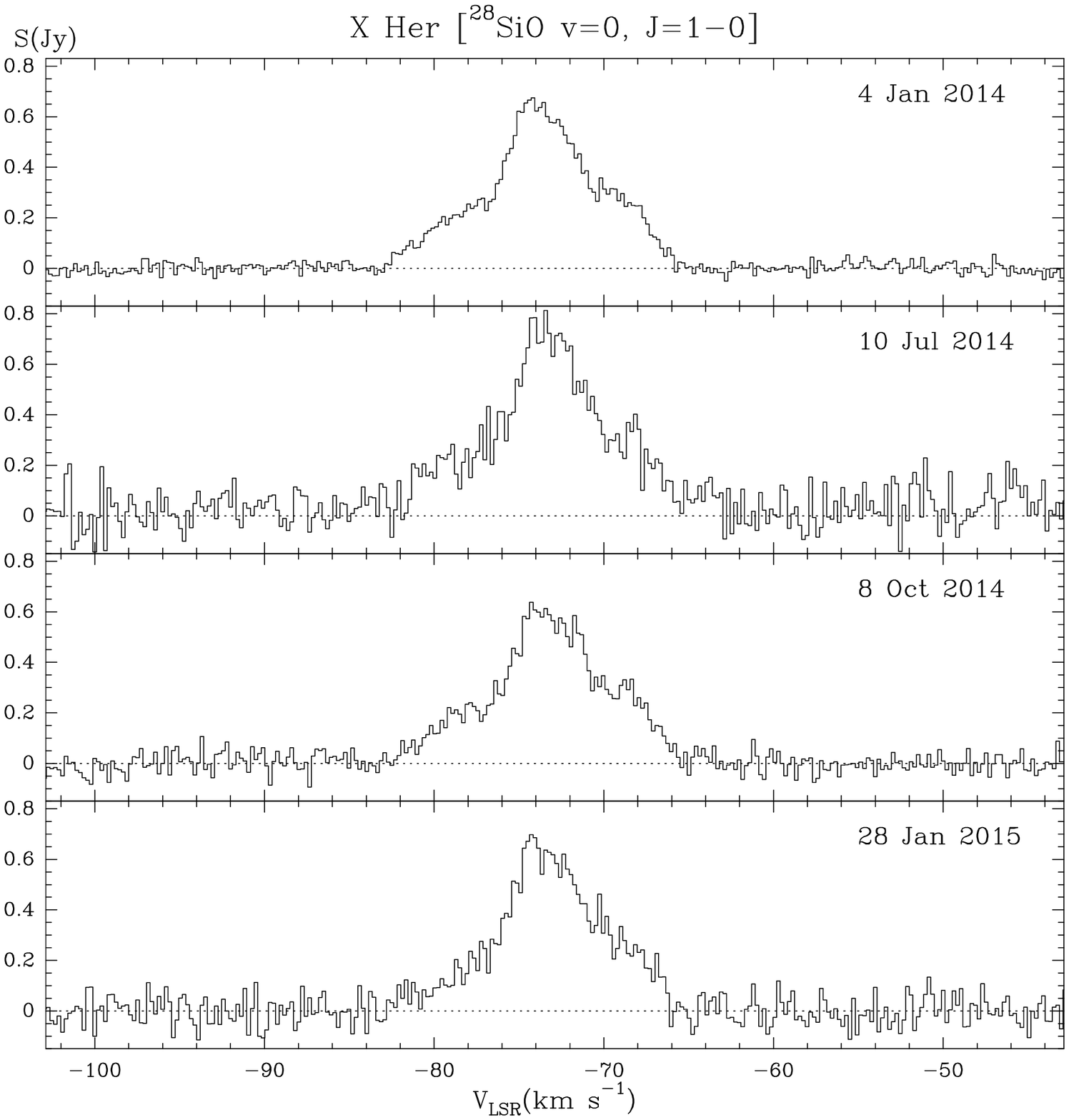}}}
   \caption{SiO $v$=0 \juc\ spectra in the S-type semiregular variable star
     X Her. The units and dates of the observations are indicated.}
    \label{f.xher}%
    \end{figure}


   \begin{figure}
   \centering{\resizebox{8cm}{!}{
   \includegraphics{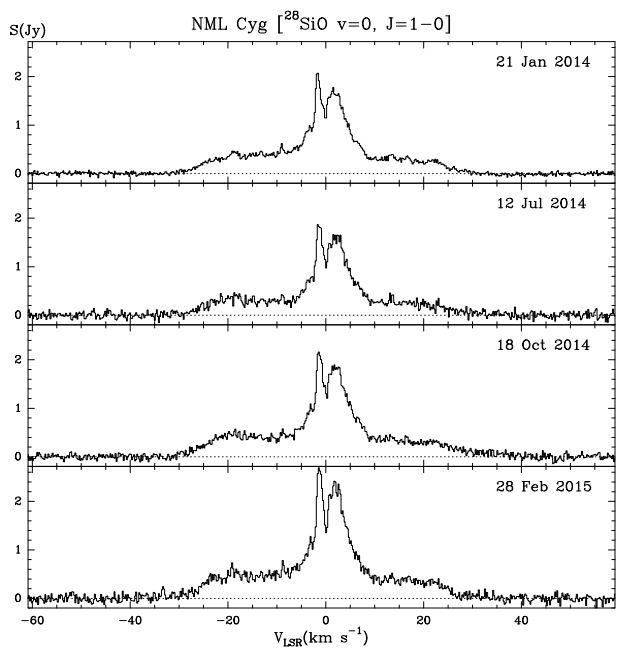}}}
   \caption{SiO $v$=0 \juc\ spectra in the red supergiant star NML
     Cyg. The units and dates of the observations are indicated.}
    \label{f.nmlcyg}%
    \end{figure}

   \begin{figure}
   \centering{\resizebox{8cm}{!}{
   \includegraphics{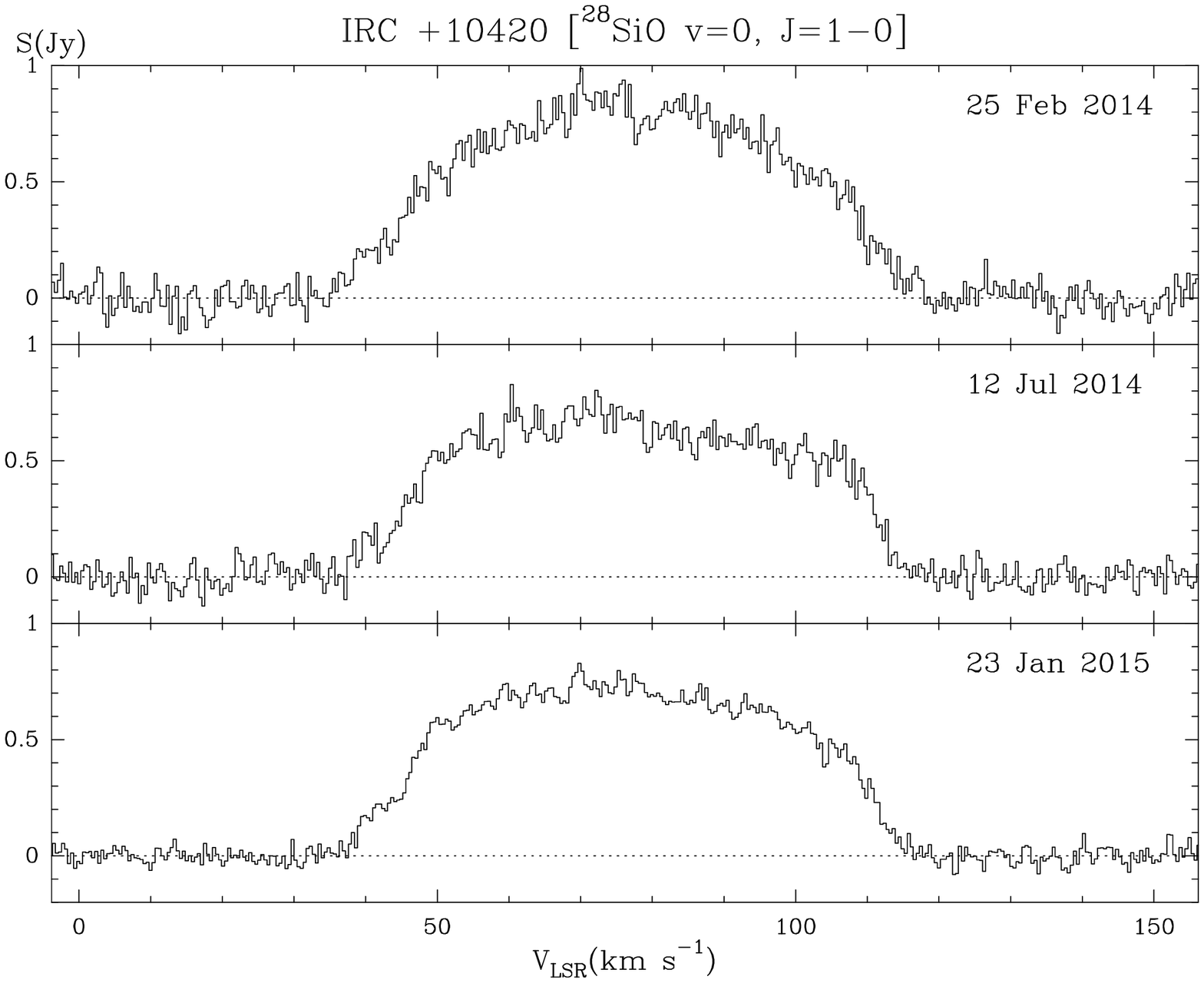}}}
   \caption{SiO $v$=0 \juc\ spectra in the yellow hypergiant star
     IRC\,10420. The units and dates of the observations are indicated.}
    \label{f.irc10420}%
    \end{figure}


\section{Observations}

The observations were performed with the 40~m radio telescope, operated by the National Geographic
Institute (IGN) in Yebes Observatory (Spain). The telescope is located at 
40$^\circ$ 31$'$ 28$''$.78 north, 03$^\circ$ 05$'$ 12$''$.71 west and 980~m above sea level. Weather conditions 
are typically dry; the amount of precipitable water toward the zenith ranges 
between 4~mm in winter to 14~mm in summer. The antenna has
an homological design and operates in Nasmyth focus. The primary
mirror backstructure is covered with a cladding and has a forced air
flow to keep the temperature homogeneous, but since it is not
hermetically isolated from the outside and temperature controlled, it
suffers some astigmatism when temperature along the day changes more
than 10 degrees. This has a relevant effect at frequencies above 70
GHz. The Half Power Beam Width (HPBW) at 43 GHz is 44 arcsecs.

The antenna is equipped with a dual-polarization (LCP and RCP) single-pixel 45
GHz cryogenic receiver mounted in a Nasmyth focus. During the observations
described in this paper the receiver was upgraded and its location moved to a
new position in the Nasmyth cabin. Observations from the first epoch, those
performed in 2013 and 2014, have a higher receiver noise temperature than those
in the second epoch. Since the beginning of 2015 the receiver was improved and
the system noise temperature dropped approximately by 30 K in both polarization
channels, to about 90 K.

Calibration at 43 GHz is carried out using the chopper wheel method. A hot
load at ambient temperature and the sky as cold load were used to
calibrate the data. Opacity was estimated theoretically using the ATM
code \citep{pardo01,pardo02} and weather conditions measured at a 
metereological station 400 m away
from the antenna. The error in the determination of the sky opacity was
estimated comparing the results from ATM with those obtained from
skydip measurements, from which the forward efficiency of the antenna
(90\%) was also estimated. In all cases the error was always below
10\%. We also tested the validity of the sky as cold load by
comparing the receiver temperature determined using this same chopper
wheel method with a hot and cold load method using liquid nitrogen. The
difference is also 10\% at most. 
All together, we estimate that the total calibration uncertainty of our
observations is $\pm$ 20\%.

The conversion between antenna temperature ($T_a^*$) and flux scales, 0.16
K/Jy, was obtained using observations toward sources with well-known 
flux and size, such as the planets Venus, Mars, Jupiter, and Saturn,
which were scheduled in all observing runs. The conversion depends
mainly on the aperture efficiency of the telescope, which we have
estimated to be 30\% $\pm$ 10$\%$ in an elevation range between 20 and
80 degrees. Systematic variations of the efficiency with
elevation were parameterized and taken into account. Small variations
(10\%) in the efficiency can be seen between epochs with very different
ambient temperatures, but this variability is included in the 
calibration error budget explained earlier.

The 45 GHz receiver has an instantaneous 8 GHz bandwidth. The signal is 
downconverted and the Intermediate Frequency (IF) split into 2 GHz bands, and later the IF split into
0.5 GHz wide bands. The bands are fed into a fast Fourier transform
spectrometer with several modules, which can be configured to provide
different band widths and spectral resolutions.
The observations described in this article used two 500 MHz modules, one per circular 
polarization, with 16384 channels per module and a spectral resolution
of 30 kHz (approximately 0.21 km/s at this frequency).

In order to keep a good pointing and focus, pseudo-continuum
observations were performed every one or two hours toward sources with
SiO $v$=1 \juc\ maser emission. Pseudo-continuum observations consist in
subtracting the summed spectral emission from the SiO maser from the rest of
the spectrum while performing a pointing drift, or a focus scan. 
This is a useful technique since local atmospheric
conditions have a small influence on the results and the baselines
obtained are rather flat. Pointing errors were always kept within 5$''$
in both axes.

Observations consisted of spatial on-off scans with an integration time
of 60 seconds, using a spatial reference 400$''$ away in azimuth 
to subtract the atmospheric emission correctly. Calibration
scans were performed every 20 minutes. The resulting baselines are good,
and only simple first order baselines were removed from the observed profiles.

The calibration of observations previous to April 2014 was only
approximate at first because before that date, the antenna efficiency
and flux-to-temperature conversion were not systematically measured
(using planets and other well-known sources). The absolute calibration
of those observations was checked by comparing the intensities of
the wide spectral components and, in particular, of the plateau (see
Sect.\ 3.1) with observations of the same sources performed later,
since these wide features are assumed to be constant. As discussed in
Sect.\ 3.2, those sources in which the plateau is identified well (IK
Tau, GX Mon, $\chi$ Cyg, etc) show significantly constant plateau
emission, at least within the calibration uncertainties. The
corrections applied to those preliminary data were always moderate,
within the general calibration uncertainty mentioned above. The narrow 
intense spikes, very clearly detected in many sources
(Sect.\ 3.1), show strong variations (as described in Sect.\ 3.2).

\section{Results}

We have observed a total of 28 stars in $^{28}$SiO $v$=0 \juc\ emission (with
rest frequency 43.423853 GHz), see
data in Figs.\ 1 to 8 and A.1 to A.20. Our sample includes O-rich,
S-type, and C-rich Mira-type variables, OH/IR stars, which are
highly-obscured O-rich Miras with strong OH maser emission, O-rich super,
and hypergiants and semiregular variables. Source properties are
summarized in Table 1 and the different kinds of observed spectra are
discussed in Sect.\ 3.1. A summary of the observational parameters is
shown in Table A.1.  Most objects were observed several times, and some
of them were more systematically monitored; see detailed discussion on
the observed variability in Sect.\ 3.2.

\subsection{General properties of SiO $v$=0 \juc\ emission from evolved stars}

The O-rich and S-type Mira-type stars systematically show composite
spectra, with a wide plateau plus more intense narrow spikes. This is
conspicuously the case of, for instance, TX Cam (Fig.\ \ref{f.txcam}), GX Mon (Fig.\ \ref{f.gxmon}), 
IK Tau (Fig.\ \ref{f.iktau}), R Leo (Fig.\ \ref{f.rleo}), and the S-type star $\chi$ Cyg
(Fig.\ \ref{f.chicyg}). The plateau occupies a spectral range similar to that of the
lines of CO, HCN, SiO $v$=0 \jdu, etc., and shares the
same centroid. The total profile width is expected to give the final expansion 
velocity, while the centroid would indicate the stellar systemic velocity, see 
discussion in Sect. 1. The spikes are very narrow and
in most cases occupy a relatively small total range also centered on the systemic
velocity. In some stars, notably in TX Cam (Fig.\ \ref{f.txcam}), R Cas (Fig.\ \ref{f.rcas}),
and probably in $\chi$ Cyg (Fig.\ \ref{f.chicyg}), spikes also appear at more extreme
velocities. The narrow spikes tend to show significant variability,
contrary to the case of the plateau, which is remarkably constant; see
discussion in Sect.\ 3.2.

$o$ Cet (Mira) is a peculiar case (Fig.\ \ref{f.omicet}). It shows a narrow
SiO $v$=0 \juc\ profile, 2--3 \kms\ wide (and some variability in the
whole spectrum). The profiles observed in CO from this source are about
twice as wide \citep{planesas90,ramstedt14} and show the same
centroid. The CO profiles are known to come from a complex shell, which is 
probably affected by the presence of a stellar companion.

We have observed three OH/IR stars. These objects, which are thought to
be very obscured O-rich Mira-type variables characterized by their intense OH maser emission, 
show profiles that are very similar
to those of standard (optically brighter) O-rich Miras. That profile
structure is particularly clear in IRC\,+10011 (Fig.\ \ref{f.irc10011}) and
OH\,44.8-2.3 (Fig.\ \ref{f.oh44}). OH\,26.5+0.6 (Fig.\ \ref{f.oh265}) seems to show
similar profiles, but with a low S/N ratio that hampers our
interpretation.

The observed SiO $v$=0 \juc\ profiles of the O-rich red supergiants NML
Cyg and VX Sgr (Figs.\ \ref{f.nmlcyg} and \ref{f.vxsgr} respectively) show properties 
that are very similar to those
of most O-rich Mira-type variables, except for the width of the 
profiles, which is wider than $\sim$ 50 \kms. In the O-rich red supergiant VY
CMa (Fig.\ \ref{f.vycma}), we find a very intense and complex line profile, with
many spikes occupying a very wide spectral region similar to the v>0 masers. 
Finally, the red
supergiant S Per (Fig.\ \ref{f.sper}) shows a complex and spiky profile, which is similar
to that of VY CMa, but the line is less intense and narrower. In all
monitored supergiant stars, the profiles are relatively constant with
time (Sect.\ 3.2), even those with narrow spikes.  The profiles observed in the
yellow hypergiant IRC\,+10420 (Fig.\ \ref{f.irc10420}) are wide and featureless, and very
similar to those observed in other molecules \citep{quintanal07}.

The C-rich Mira-type variables IRC\,+10216 (CW Leo, Fig.\ \ref{f.irc10216}), LP And
(Fig.\ \ref{f.lpand}) and CIT\,6 (RW LMi, Fig.\ \ref{f.cit6}) show wide profiles with
the standard shapes that are often found in AGB CSEs (and, in particular, in
C-rich stars). However, the SiO $v$=0 \juc\ line is weak in these Mira-type variables
compared with emission from other sources and with lines of C-bearing
molecules observed in these sources.

We observed six O-rich semiregular variables, three of which, 
RX Boo, RS Cnc, and X Her, are known to present axisymmetric 
circumstellar shells; see Sect.\ 1. In all six cases, the SiO $v$=0 \juc\ profiles are
comparable to those observed in CO. In X Her (Fig.\ \ref{f.xher}), the SiO $v$=0
\juc\ profile is composite, which is similar to those we obtained in O-rich
Miras, but with no sign of the spiky structure in the central spectral
feature found in Miras. In this source, the composite CO profile is
known to be due to the axial symmetry observed in interferometric maps
\citep{ccarrizo10}, and the spectral central component comes from a very
slow expanding disk or torus. RX Boo (Fig.\ \ref{f.rxboo}) shows axial symmetry in
CO maps, but the single-dish CO profiles are simple and very
similar to our featureless SiO $v$=0 \juc\ line. RS Cnc (Fig.\ \ref{f.rscnc}) shows CO
spectra and shell structure similar to those of X Her; our data
also show composite profiles similar to those of CO and to
those found in X Her. In two other semiregular variables, R Crt and RT
Vir (Figs.\ \ref{f.rcrt} and \ref{f.rtvir}), the SiO $v$=0 \juc\ and CO profiles are
also wide and mostly featureless; no evidence of axial symmetry has
been found so far in these sources. Finally, W Hya is often catalogued
as a semiregular variable, but its intensity curves are relatively
regular both in the optical and SiO maser emission \citep{pardo04}. The SiO
$v$=0 \juc\ profiles in W Hya (Fig.\ \ref{f.whya}) vaguely show a composite
structure that is similar to that of O rich Miras which could be related 
to the fact that W Hya is considered almost a regular variable in 
some aspects.

\subsection{Time variability of the $v$=0 \juc\ SiO profiles}

In order to check the time variability of the SiO $v$=0 \juc\ emission,
line profiles were obtained several times in a number of stars, during a
time period of longer than one year between about November 2013 and April
2015. We systematically monitored the stars TX Cam, R Cas, $o$ Cet, IK
Tau, $\chi$ Cyg, IRC\,+10216, VY CMa, NML Cyg, and X Her. As many as six 
independent observations were obtained in the best cases; other sources 
were observed two or three times.

As mentioned, the profiles in most O-rich and S-type Mira variables are
composite with a wide plateau plus narrow spikes that often occupy the
central parts of the profiles. The wide plateau is remarkably constant,
within the uncertainties (Sect.\ 2). We can see this clearly in the
cases of IK Tau, TX Cam, and the S-type star $\chi$ Cyg. The spikes,
however, significantly vary, showing clear changes in their relative
intensities. In any case, the overall variations of the SiO $v$=0
\juc\ is much more moderate than in the very strong $v$$\geq$1 masers;
see for example, \cite{pardo04}. Those strong maser lines can vary by orders of
magnitude in both line peak and area, while these parameters rarely
vary by a factor of two in our $v$=0 \juc\ observations (see, however, the
exceptional cases of $\chi$ Cyg and R Cas in Figs. \ref{f.chicyg} and \ref{f.rcas}, respectively).

In two cases, namely TX Cam (Fig.\ \ref{f.txcam}) and R Cas (Fig.\ \ref{f.rcas}), some spikes appear at more
extreme velocities but not in all epochs. The variability of these
satellite peaks is, in the few observed cases, significantly
higher than for the rest of the observed features. In the case of
$\chi$ Cyg (Fig.\ \ref{f.chicyg}), there is a relative maximum at about 15 \kms\ $LSR$,
which could be one of these moderate velocity spikes. However its low intensity
and lack of significant variability (at our detection levels)
prevent any conclusion in this respect. The variability of the
particularly narrow profile found in $o$ Cet (Fig.\ \ref{f.omicet}) is noticeable but
moderate, and we can see changes in the profile shape.

In general, the variability of the spikes seems to be chaotic, without
any obvious dependence on the optical or IR phase. Only the
moderate velocity spike in R Cas could be associated with the phase,
appearing more or less at the same time as the optical maximum of
2014. However, the moderate velocity peaks seen in TX Cam, between 20
and 25 \kms, are not correlated with the optical phase. We conclude
that there is no clear association between the optical or IR cycle and the
variations found in the $v$=0 \juc\ spikes.

Some time variability is observed in the red supergiants, but this vairability 
is weaker than in the previously mentioned objects and much weaker than the very
large variations found in their strong $v$$>$0 maser emission
\citep{pardo04}. In these objects the structure of the profiles is kept
significantly constant with time.  The SiO $v$=0 \juc\ emission from
other sources observed several times (semiregular variables and C-rich
Miras) do not show any clear variability. 
The profiles of the O-rich semiregular
variable X Her, which are composite and similar to those of O-rich
Miras (except for the spiky structure of the Miras), are remarkably
constant within the uncertainties, see Fig.\ \ref{f.xher}.

\section{The origin of the two-component profiles of the $^{28}$SiO
  $v$=0 \juc\ line}

The relatively intense spikes that form the central spectral feature
systematically observed in SiO $v$=0 \juc\ emission from O-rich and
Miras and the S-type Mira $\chi$ Cyg are very narrow and time variable,
suggesting that they are the result of maser emission
\citep{bobcla04}. Their intensity is higher than that of the
surrounding plateau; but it is much weaker, typically a hundred times
weaker in flux units, than the intensity of the SiO $v$$\geq$1 masers
observed in these sources. We recall that the vibrationally excited
levels are expected to be strongly underpopulated. The energy of the
$v$=1 levels is about 1800~K higher in energy than the ground $v$=0
levels (and the $v$=2 levels are at about 3600 K from the
ground). Moreover, the vibrational excitation temperature is expected to be of just
some hundred Kelvin, even for high kinetic temperatures, since the very
fast spontaneous decay of the vibrationally excited levels (faster than
5 s$^{-1}$) prevents thermalization \citep[in fact the $v$$>$0
  levels are mostly populated by the stellar IR radiation, see for example][]{bujetal96}
Therefore, the relative population of the vibrationally excited
rotational levels must be smaller than 10\%.

From their VLA maps, \cite{bobcla04} conclude that these spikes come
from relatively wide spots, the strongest spike of each source is about
0\secp 2-0\secp 4 wide, in which the brightness is moderately high,
10$^4$-10$^5$ K. Meanwhile, the $v$$>$0 maser spots are really tiny, at a
few mas wide, and show a very high brightness \gsim\ 10$^9$ K
\citep{diamond94,desmurs14}. In both cases, the maser spots are placed
at a few stellar radii from the star, and the $v$=0 \juc\ emission is somewhat 
more extended; in angular units, the total extent is typically
found to be $\sim$ 0\secp 2 for $v$=0 \juc\ emission and \lsim\ 
0\secp 1 for $v$$>$0 masers.

The only way to explain the low brightness of the $v$=0 \juc\ spikes
relative to the $v$$>$0 masers, is to assume that the amplification of
the maser effect responsible for the $v$=0 \juc\ (weak) masers is much
lower than that of the strong masers. From the above mentioned
brightness levels, we expect $|\tau$($v$=0~\juc)$|$ $\sim$
$|\tau$($v$=1~\juc)$|$ -- 10; in a rough estimate, $|\tau$($v$=0 \juc)$|$
must be of just a few units.

It is surprising that, in spite of the much higher population of the
$v$=0 rotational levels, the opacity modulus is so low, particularly
compared with those of the $v$$>$0 masers. We recall that $\tau$
\aprop\ $x/T_{\rm ex}$, where $x$ is the relative population per
magnetic sublevel (i.e., the average of the upper and lower levels,
$x$=$(n_u/g_u+n_l/g_l)/2)$ and $T_{\rm ex}$ is the excitation
temperature of the maser transition. Therefore, $T_{\rm ex}$
(\aprop\ $\frac{x_u}{x_l-x_u}$) must be very high in absolute value,
i.e.\ the population inversion must be extremely low with $x_u$
practically identical to $x_l$ and $x_u-x_l$ $\ll$ $x$. The inversion
of the $v$$>$0 masers is already not very high in the intense spots,
with $T_{\rm ex}$ $\sim$ $-$100 K (since the maser is probably
saturated). In the $v$=1 \juc\ maser, for instance, the
inversion is just of about 2\%, $x$($J$=1) $\sim$
$x$($J$=0)$\times$1.02. Hence we find a conservative upper limit of 0.1\% on 
the $v$=0 \juc\ population inversion in the maser clumps that would be 
responsible for the observed narrow spikes in our profiles. Such a fine
tuning of the level population leading to a very slight population
inversion seems improbable and able to operate only in very peculiar
conditions, but the $v$=0 \juc\ spikes are ubiquitous in O-rich
Mira-type stars. The inversion of this transition would be due to an
inefficient pumping mechanism, leading to very easily saturated maser
emission with relatively low brightness levels. This is a situation that, to
our knowledge, has only been found in this case. We think that the actual origin 
of the $v$=0 \juc\ spikes is not well understood yet.

\subsection{Spectral features in O-rich Mira-type and semiregular variables}

The profile structures found in O-rich Mira-type variables (including
the S-type star $\chi$ Cyg) and in the semiregular variables X Her and
RS Cnc are remarkable. In all cases, the line is composite, with a
central stronger feature plus a wider plateau. The central feature is,
however, different. While in Miras variables that component is composed of a
number of narrow time-variable spikes, in the semiregulars 
it is smooth and does not show significant variability in our
observations; this is probably because of thermal emission instead of
population inversion. The plateau is in all cases constant with time and
seems to be due to thermal emission.

The origin of the composite structure of the line profiles in X Her and
RS Cnc (found not only in $v$=0 \juc, but also in the other molecular
lines, Sect.\ 1) is known to be due to the overall CSE structure, which
is also complex. In contrast to what happens in
most Miras, whose shells show a significant spherical symmetry
at scales larger than 1$''$, (Sect.\ 1) the CSEs around these
semiregular variables show an equatorial slowly expanding disk and a
bipolar low-collimation outflow, which are responsible for
the central component and the plateau of 
the lines of SiO (and other molecules), respectively. This shell
structure is similar, although with lower velocities, to
that commonly observed in protoplanetary nebulae, which is the obvious
precedent of the strong axial symmetry found in planetary nebulae
(often showing torus-like or double-bubble structures). 

As we have seen, we also find in Mira-type O-rich stars two-component
profiles, in which the central spikes seem to show weak maser
effects. The simplest explanation of these composite profiles is to
assume that they come from the superposition of a standard wide
profile, due to {\it thermal} emission from expanding shells, and a
group of weak maser spikes, whose low projected velocity would be due
to a predominantly tangential amplification (as found in the very
intense $v$$>$0 masers, Sect.\ 1). As an alternative, we speculate that
the similar $v$=0 \juc\ profile shapes shown by these objects and by
the above mentioned semiregular variables could indicate that,
independent of the maser or thermal nature of the
emission, the line shapes in both kinds of objects have a common
origin: the presence of axial symmetry in the emitting region. 
In O-rich Miras, such an asymmetry would only appear in the very
compact SiO emitting region, and at large scale would in some way disappear or 
be more difficult to detect, but would become dominant in later evolutionary phases.
We finally note that clear composite structures are not usually found
in higher-$J$ $v$=0 SiO profiles (Sect.\ 1). The reason for this could be that the
maser effect found in $v$=0 \juc, which is certainly very peculiar,
does not appear in other $v$=0 transitions and/or that absorption by
outer regions is very high in those lines; the opacity of high-$J$
lines is expected to be high, since it increases roughly
proportionally to $J^2$ under these conditions of high excitation.
The lack of maps
with high sensitivity (allowing good images of the whole profiles) and
angular resolution (allowing images at 0\secp 1 scales) prevents
any firm conclusion at the moment. We hope that future high-quality
observations will throw light on this topic.

The two-component profiles also found in (at least some) red
supergiants could also indicate that phenomena similar to those described 
above are present in their inner
shells. The SiO $v$=0 profiles of C-rich AGB stars (and some S-type objects)
are, however, smooth and wide, which is very similar to those observed in
the emission of other molecules. Maps of $v$=0 \jdu\ emission (Sect.\ 1)
show that its extent is significantly larger than for O-rich Miras, which 
is attributed to the different grain formation processes, but with 
a relatively low SiO abundance.  
The difference in the $v$=0 profiles of C-rich and O-rich Mira-type stars
would be due to these properties. 

\section{Summary and conclusions}


We present observations of SiO $v$=0 \juc\ emission from 28 evolved
stars, including O-rich, S-type and C-rich Mira-type variables, OH/IR
(O-rich) stars, O-rich semiregular long-period variables, red supergiant
stars, and one yellow hypergiant star.  All these stars are more or
less regular pulsators with long periods of about one year or more and are known
to be surrounded by thick circumstellar envelopes (CSEs) emitting in
maser and thermal (not masing) molecular lines; see Sect.\ 1. We
performed studies of the emission variability in about one half of the
sources, spanning over more than one year. 

In most O-rich and S-type Mira-type stars (including the OH/IR
objects), the observed profiles are composite, with a wide spectral
feature or plateau and a set of narrow, relatively intense spikes 
in the line center (Sect.\ 3.1). 
The plateau occupies roughly the same velocity range as the
smooth profiles characteristic of thermal molecular lines 
(CO, HCN, SiO $v$=0 \jdu, etc). The spikes occupy a central narrower
spectral region and are probably the result of weak maser effects (see below).
A peculiar case is $o$ Cet, (Mira, a binary O-rich star showing an asymmetric shell), which 
shows narrow spiky profiles of SiO $v$=0 \juc.

The SiO $v$=0 \juc\ line in C-rich stars shows smooth featureless
profiles, very similar to those of the other molecules and probably due
to thermal emission.

The observed supergiants also show complex profiles with many
spikes. NML Cyg, S Per, and VX Sgr show two-component profiles similar
to those observed in O-rich Miras (except for the larger total
width). The yellow hypergiant IRC\,+10420 shows a 
smooth profile, which is very similar to those of standard lines.

The SiO $v$=0 \juc\ profiles of the selected semiregular giants are
smooth, except for the spiky profiles found in W Hya (a source that
shows a very regular variability and is sometimes catalogued as
Mira-type variable).  Two semiregulars, X Her and RS Cnc, show
composite profiles including a wide plateau plus a central stronger
feature, but this profile is not composed of narrow spikes.
Remarkably, the profiles of standard lines observed
in X Her and RS Cnc, e.g., CO \juc\ and \jdu\ (see Sects.\ 1, 3.1), are
practically identical to those we find and 
are known to be associated with a significant
axial symmetry of the circumstellar envelope.

Significant variability was found in the narrow spikes we detect in SiO
$v$=0 \juc\ emission, mostly in the O-rich regular pulsators and the
S-type Mira-type star $\chi$ Cyg (Sect.\ 3.2). However, the wide
features, including the plateaus clearly identified in many sources,
the two spectral components observed in X Her and RS Cnc, and the whole
profiles of C-rich objects, are significantly constant within the
calibration uncertainties. The variations of the
narrow spikes are not clearly correlated with the stellar pulsation cycle.

The relatively intense spikes that form the central spectral feature
systematically observed in O-rich and Miras and the S-type Mira $\chi$
Cyg are very narrow and time variable, suggesting that they are the
result of maser emission (Sect.\ 3). Their intensity is higher than
that of the surrounding plateau, but it is much lower, typically by a
hundred times, than the intensity of the SiO $v$$>$0 masers observed in
these sources.
The low flux and brightness characteristic of these spikes,
despite the high relative population of the $v$=0 rotational levels
(Sect.\ 4), indicates that probably the maser pumping of the $v$=0
\juc\ spikes is very peculiar
and much less efficient than that of the $v$$>$0 masers.

The simplest explanation of the two-component $v$=0 \juc\ profiles
observed in O-rich and S-type regular pulsators is that they are due to
the superposition of standard wide profiles and a group of weak maser
spikes coming from very inner layers, for which, as happens for the
$v$$>$0 strong masers, tangential amplification leads to a
relatively narrow total dispersion in velocity; see Sect 4.1. 
We speculate that it would be
useful, if confirmed, to study the onset of axial structures in the post-AGB 
evolution, and that the similar
profiles found in O-rich and S-type regular pulsators and in 
semiregular variables may have a similar origin (regardless of possible maser effects in the narrow spikes), indicating the systematic
presence of axial symmetry in the innermost layers of most AGB
CSEs. The CSEs of O-rich and S-type Miras are known to be spherical and
isotropical at scales \gsim\ 1$''$; their SiO thermal emission comes
precisely from regions $\sim$ 1$''$ wide. Future high-quality maps at this 
frequency are expected to solve this question.


\begin{acknowledgements}
      This work is based on observations carried out with the  
      IGN 40~m radio telescope. We thank all the staff in Yebes Observatory for 
      the generous help received. We acknowledge partial 
      funding from MINECO grants AYA2012-32032, FIS2012-32096, and 
      FIS2012-38160. This work has
      made use of the American Association of Variable Star Observers
      (AAVSO) database and of the SIMBAD Astronomical database
      (operated at CDS, Strasbourg, France). 
\end{acknowledgements}


{}

~

\newpage

~

\appendix

\section{Further observational data of $v$=0 \juc}

This Appendix contains the spectra of stars not shown previously and the table summarizing the observational results.

\begin{table}
\caption{Summary of observational results. *: data with 1$^{st}$ receiver.
}  
\scriptsize
\vspace{-.2cm}
\begin{tabular}{|lccccc|}
\hline
source & date &  area & peak  & $\sigma$  &
$V_{LSR} (centroid)$  \\ 
 & & Jy\,\kms & Jy & Jy & \kms \\
\hline
\hline
\multirow{2}{*}{RR Aql} & Jul-Aug 2014* & 3.9 & 0.66 & 0.03 & 28.1 \\
 & 2 Mar 2015 & 2.6 & 0.51 & 0.07 & 28.5 \\
\hline
\multirow{6}{*}{TX Cam} & 27 Nov 2013* & 30.8 & 1.80 & 0.016 & 10.0 \\
 & 13 Jul 2014* & 29.4 & 1.88 & 0.09 & 11.1 \\
 & 2 Oct 2014* & 34.5 & 1.98 & 0.08 & 10.8 \\
 & 20 Oct 2014* & 35.2 & 2.06 & 0.07 & 11.3 \\
 & 28 Jan 2015 & 38.9 & 1.99 & 0.07 & 11.0 \\
 & 2 Mar 2015 & 30.7 & 1.79 & 0.05 & 10.6 \\
\hline
\multirow{6}{*}{R Cas} & 23 Feb 2014* & 20.4 & 2.15 & 0.03 & 25.8 \\
 & 16 Mar 2014* & 21.6 & 2.8 & 0.1 & 25.6 \\
 & 5 Aug 2014* & 26.8 & 3.46 & 0.05 & 25.3 \\
 & 20 Oct 2014* & 29.1 & 3.74 & 0.05 & 25.4 \\
 & 27 Jan 2015 & 45.5 & 6.02 & 0.12 & 25.4 \\
 & 28 Feb 2015 & 48.7 & 9.39 & 0.12 & 25.2 \\
\hline
\multirow{6}{*}{$o$ Cet} & 21 Jan 2014* & 4.7 & 2.93 & 0.16 & 46.1 \\
 & 24 Feb 2014* & 3.0 & 2.07 & 0.15 & 46.3 \\
 & 10 Jul 2014* & 5.5 & 3.57 & 0.06 & 46.4 \\
 & 23 Oct 2014* & 5.2 & 4.45 & 0.11 & 46.7 \\
 & 27 Jan 2014* & 4.8 & 2.72 & 0.08 & 46.6 \\
 & 28 Feb 2014* & 4.7 & 2.3 & 0.1 & 46.6 \\
\hline
\multirow{1}{*}{R Hya} & Jul-Aug 2014* & 3.8 & 0.56 & 0.06 & -10.4 \\
\hline
\multirow{3}{*}{R Leo} & 11 Jan 2014* & 12.5 & 2.7 & 0.05 & 0.1 \\
 & 11 Jul 2014* & 18.7 & 5.86 & 0.08 & 0.0 \\
 & 1 Mar 2015 & 22.8 & 7.17 & 0.11 & 0.0 \\
 & 16 Apr 2015 & 21.6 & 8.0 & 0.5 & 0.0 \\
\hline
\multirow{4}{*}{R LMi} & 8 Mar 2014* & 5.7 & 0.69 & 0.06 & 0.3 \\
 & 17 Jul 2014* & 2.9 & 0.55 & 0.08 & 1.0 \\
 & 28 Jan 2015 & 4.4 & 0.62 & 0.05 & 0.4 \\
 & 1 Mar 2015 & 5.6 & 0.93 & 0.09 & 5.6 \\
\hline
\multirow{3}{*}{GX Mon} & 17 Jul 2014* & 24.8 & 1.85 & 0.07 & -10.4 \\
 & 6 Aug 2014* & 25.3 & 2.11 & 0.08 & -10.3 \\
 & 1 Mar 2015 & 28.4 & 1.66 & 0.11 & -10.4 \\
\hline
\multirow{6}{*}{IK Tau} & 30 Dec 2013* & 64.0 & 5.66 & 0.03 & 33.9 \\
 & 30 Jul 2014* & 63.0 & 4.89 & 0.09 & 33.6 \\
 & 2 Oct 2014* & 57.6 & 4.26 & 0.07 & 33.8 \\
 & 22 Oct 2014* & 61.0 & 4.64 & 0.08 & 34.0 \\
 & 28 Jan 2015 & 62.0 & 5.09 & 0.07 & 33.6 \\
 & 28 Feb 2015 & 65.2 & 6.0 & 0.1 & 33.7 \\
\hline
\multirow{1}{*}{W Aql} & Jul-Aug 2014* & 6.6 & 0.36 & 0.04 & -27.2 \\
\hline
\multirow{4}{*}{$\chi$ Cyg} & 28 Nov 2013* & 20.0 & 3.11 & 0.03 & 10.0 \\
 & 12 Jul 2014* & 29.7 & 9.03 & 0.06 & 9.9 \\
 & 19 Oct 2014* & 22.6 & 4.86 & 0.07 & 9.6 \\
 & 3 Mar 2015 & 23.6 & 7.05 & 0.07 & 9.5 \\
\hline
\multirow{3}{*}{IRC +10011} & 16 Mar 2014* & 40.0 & 2.53 & 0.13 & 9.9 \\
 & 11 Jul 2014* & 38.0 & 2.34 & 0.08 & 9.5 \\
 & 1 Mar 2015 & 28.2 & 1.70 & 0.09 & 9.2 \\
\hline
\multirow{2}{*}{OH 26.5+0.6} & 9 Mar 2014* & 5.6 & 0.6 & 0.1 & 29.7 \\
 & 10 Jul 2014* & 7.0 & 1.11 & 0.06 & 29.0 \\
\hline
\multirow{1}{*}{OH 44.8-2.3} & Jul-Aug 2014* & 6.3 & 0.667 & 0.024 & -71.2 \\
\hline
\multirow{1}{*}{LP And} & Aug 2014* & 1.9 & 0.104 & 0.017 & -18.5 \\
\hline
\multirow{1}{*}{CIT 6} & Aug 2014* & 3.4 & 0.230 & 0.022 & 4.3 \\
\hline
\multirow{4}{*}{IRC +10216} & 25 Feb 2014* & 44.4 & 2.11 & 0.06 & -26.2 \\
 & 11 Jul 2014* & 44.2 & 2.05 & 0.09 & -26.4 \\
 & 14 Oct 2014* & 41.4 & 2.34 & 0.17 & -26.0 \\
 & 25 Jan 2015 & 54.4 & 2.5 & 0.1 & -26.3 \\
\hline
\multirow{2}{*}{RX Boo} & 23 Feb 2014 & 16.6 & 1.37 & 0.03 & 1.0 \\
 & 12 Jul 2014 & 16.9 & 1.46 & 0.11 & 1.2 \\
\hline
\multirow{1}{*}{RS Cnc} & Jul-Aug 2014* & 2.8 & 0.4 & 0.03 & 7.4 \\
\hline
\multirow{2}{*}{R Crt} & Jul-Aug 2014* & 13.4 & 0.89 & 0.05 & 11.3 \\
 & 3 Mar 2015 & 13.3 & 1.0 & 0.1 & 11.3 \\
\hline
\multirow{4}{*}{X Her} & 4 Jan 2014* & 5.0 & 0.66 & 0.02 & -73.7 \\
 & 10 Jul 2014* & 5.7 & 0.81 & 0.07 & -73.02 \\
 & 8 Oct 2014* & 5.1 & 0.63 & 0.04 & -73.2 \\
 & 28 Jan 2015 & 4.6 & 0.70 & 0.05 & -72.8 \\
\hline
\multirow{3}{*}{W Hya} & 16 Mar 2014* & 16.5 & 2.63 & 0.08 & 39.9 \\
 & 3 Mar 2015 & 16.7 & 4.09 & 0.17 & 38.9 \\
 & 23 Apr 2015 & 17.2 & 4.75 & 0.15 & 38.2 \\
\hline
\multirow{1}{*}{RT Vir} & Jul 2014 & 5.9 & 0.63 & 0.06 & 17.9 \\
\hline
\multirow{5}{*}{VY CMa} & 12 Jan 2014* & 356. & 44.5 & 0.14 & 17.2 \\
 & 12 Jul 2014* & 340. & 44.4 & 0.3 & 16.9 \\
 & 20 Oct 2014* & 323. & 41.6 & 0.17 & 17.3 \\
 & 26 Jan 2015 & 295. & 37.7 & 0.3 & 16.9 \\
 & 28 Feb 2015 & 272. & 29.6 & 0.4 & 16.5 \\
\hline
\multirow{4}{*}{NML Cyg} & 21 Jan 2014* & 28.1 & 2.07 & 0.03 & -0.8 \\
 & 12 Jul 2014* & 26.4 & 1.89 & 0.06 & -2.0 \\
 & 18 Oct 2014* & 27.5 & 2.11 & 0.06 & 0.0 \\
 & 28 Feb 2015 & 36.7 & 2.71 & 0.07 & -0.9 \\
\hline
\multirow{1}{*}{S Per} & Jul-Aug 2014* & 6.7 & 0.41 & 0.03 & -41.1 \\
\hline
\multirow{2}{*}{VX Sgr} & Jul-Aug 2014* & 25.3 & 2.64 & 0.05 & 5.3 \\
 & 3 Mar 2015 & 18.0 & 2.22 & 0.14 & 5.5 \\
\hline
\multirow{3}{*}{IRC +10420} & 25 Feb 2014* & 46.6 & 1.05 & 0.09 & 76.6 \\
 & 12 Jul 2014* & 40.3 & 0.83 & 0.08 & 77.1 \\
 & 23 Jan 2015 & 58.6 & 1.24 & 0.09 & 76.2 \\
\hline
\end{tabular}
\label{t.obsresults}
\end{table}

   \begin{figure}
   \centering{\resizebox{8cm}{!}{
   \includegraphics{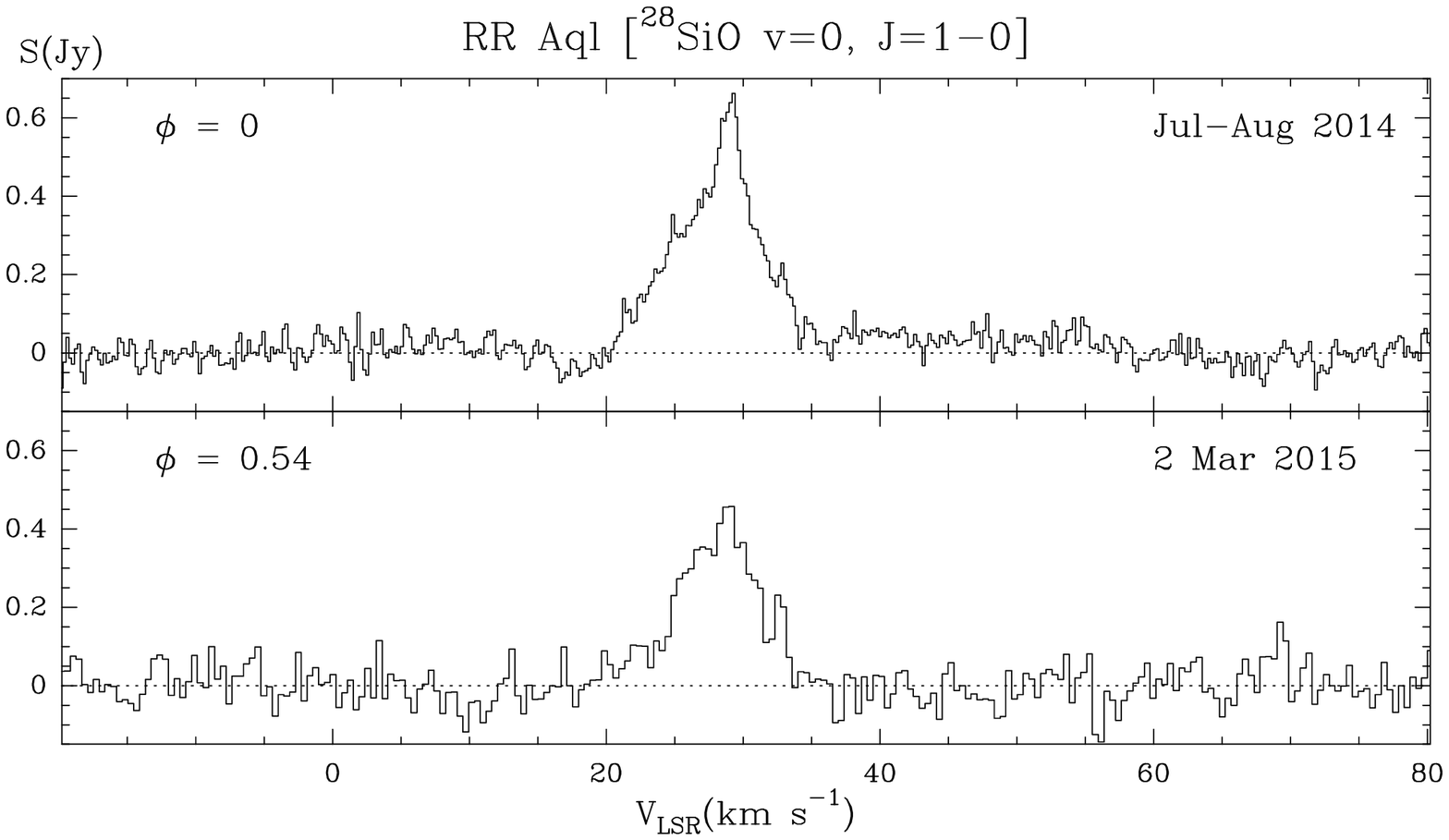}}}
   \caption{SiO $v$=0 \juc\ spectra in the O-rich Mira-type variable star
     RR Aql. The units and dates of the observations are indicated.
     The phase in the visible is shown in the upper left corner.}
    \label{f.rraql}%
    \end{figure}

   \begin{figure}
   \centering{\resizebox{8cm}{!}{
   \includegraphics{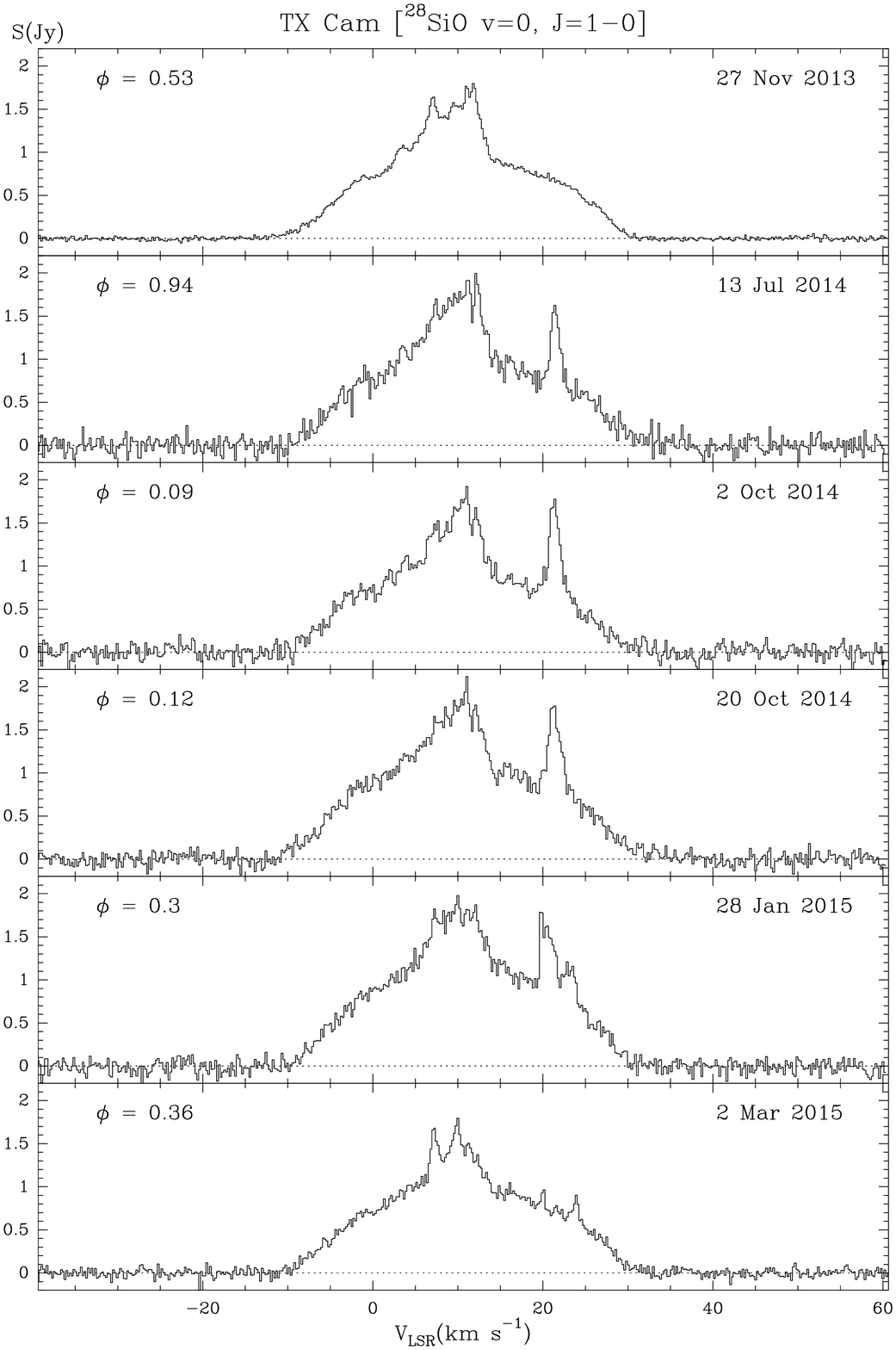}}}
   \caption{SiO $v$=0 \juc\ spectra in the O-rich Mira-type variable star
     TX Cam. The units and dates of the observations are
     indicated. The phase in the visible is shown in the upper left corner.  }
    \label{f.txcam}%
    \end{figure}

   \begin{figure}
   \centering{\resizebox{8cm}{!}{
   \includegraphics{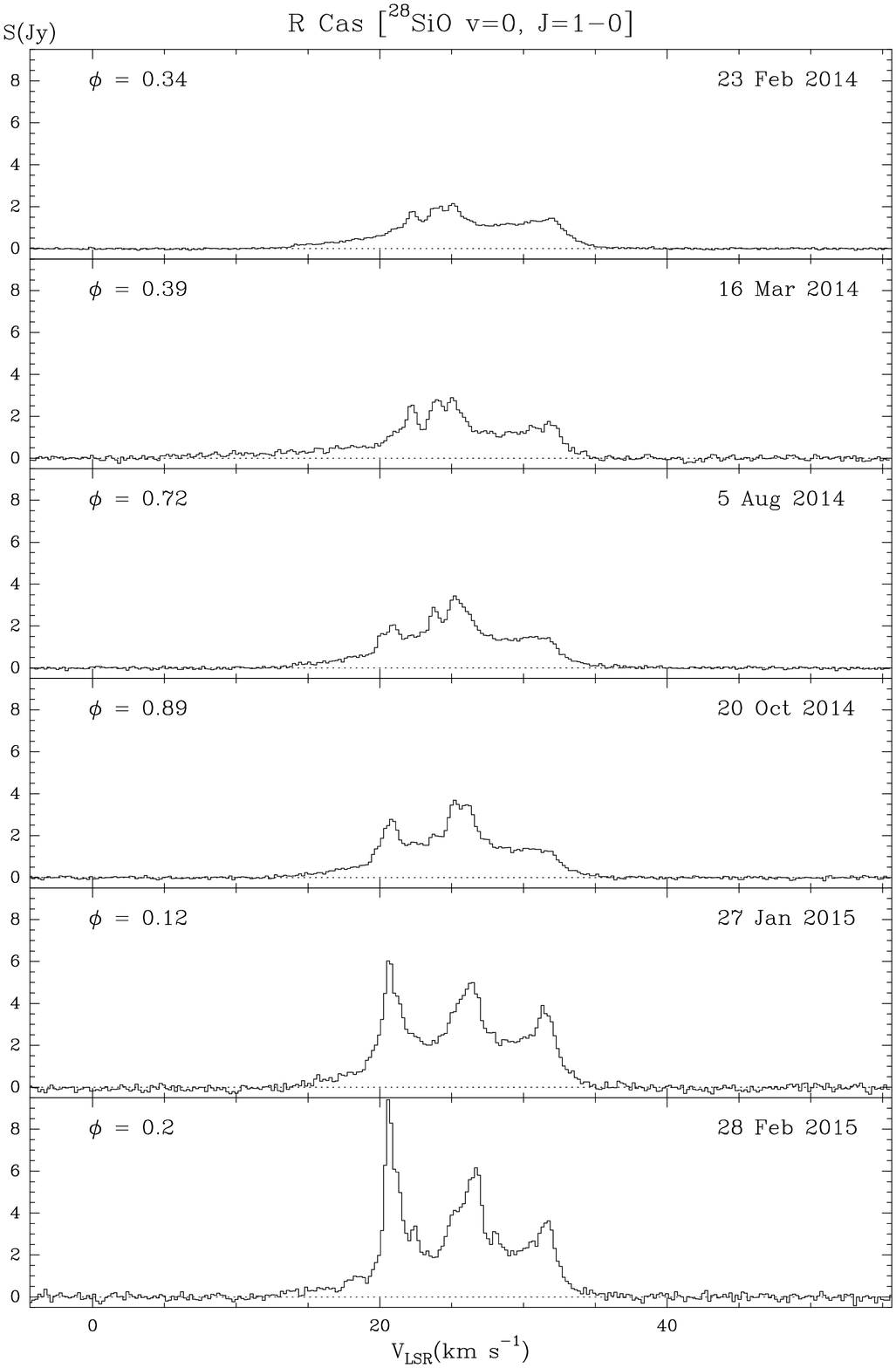}}}
   \caption{SiO $v$=0 \juc\ spectra in the O-rich Mira-type variable star R
     Cas. The units and dates of the observations are indicated. The phase 
     in the visible is in the upper left corner.}
    \label{f.rcas}%
    \end{figure}

   \begin{figure}
   \centering{\resizebox{8cm}{!}{
   \includegraphics{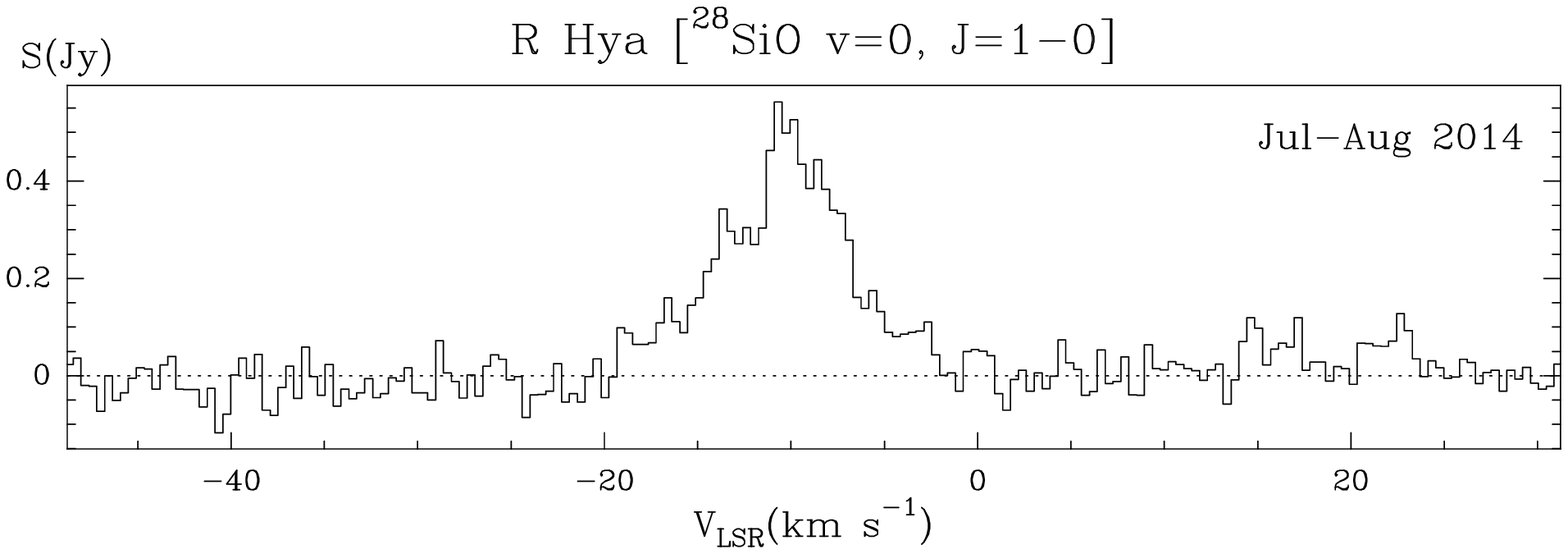}}}
   \caption{SiO $v$=0 \juc\ spectra in the O-rich Mira-type variable star R
     Hya. The units and dates of the observations are indicated.}
   \label{f.rhya}%
    \end{figure}


   \begin{figure}
   \centering{\resizebox{8cm}{!}{
   \includegraphics{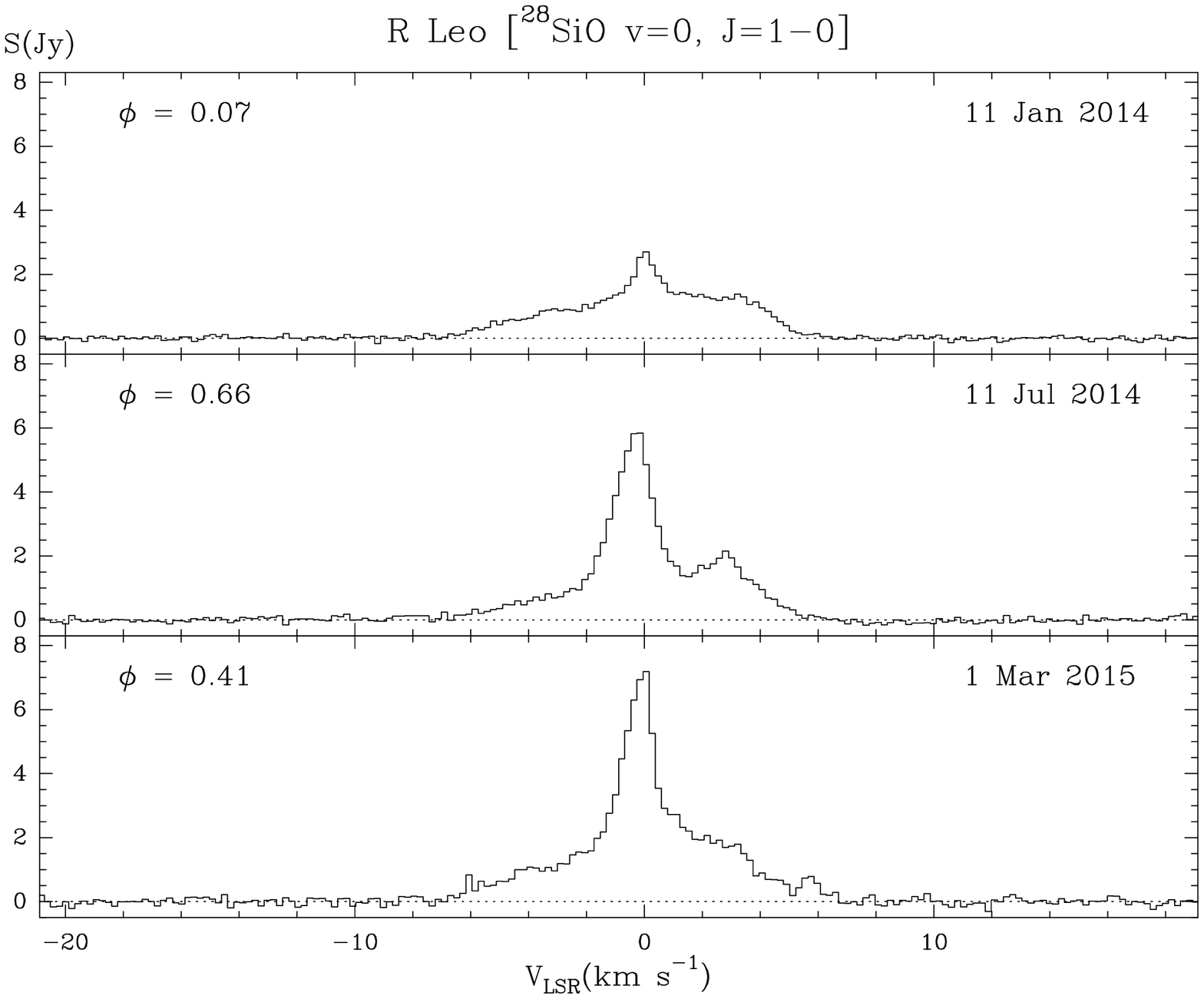}}}
   \caption{SiO $v$=0 \juc\ spectra in the O-rich Mira-type variable star R
     Leo. The units and dates of the observations are indicated.
     The phase in the visible is shown in the upper left corner.}
    \label{f.rleo}%
    \end{figure}

   \begin{figure}
   \centering{\resizebox{8cm}{!}{
   \includegraphics{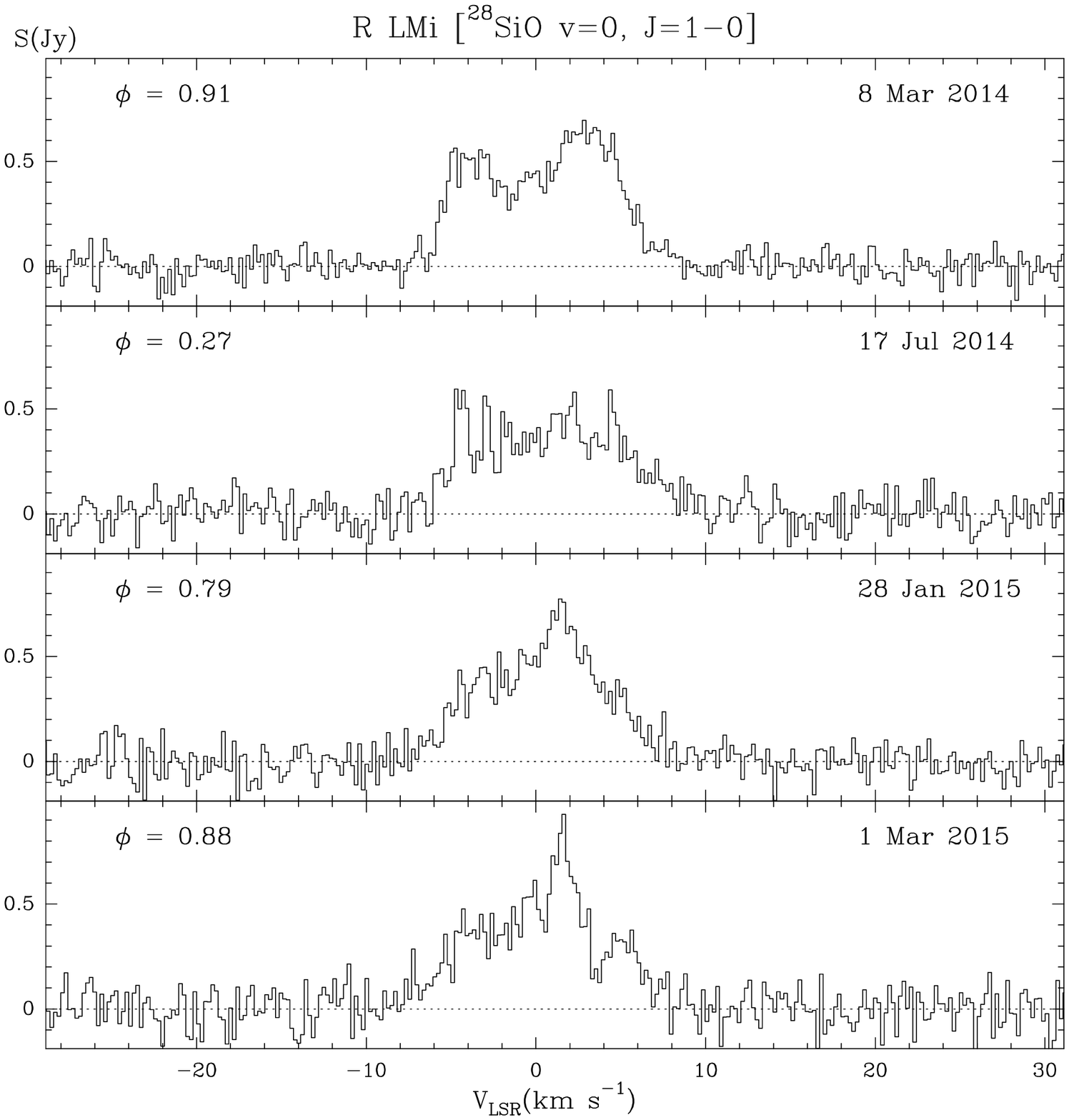}}}
   \caption{SiO $v$=0 \juc\ spectra in the O-rich Mira-type variable star R
     LMi. The units and dates of the observations are indicated. The phase 
   in the visible is in the upper left corner.}
   \label{f.rlmi}%
    \end{figure}

   \begin{figure}
   \centering{\resizebox{8cm}{!}{
   \includegraphics{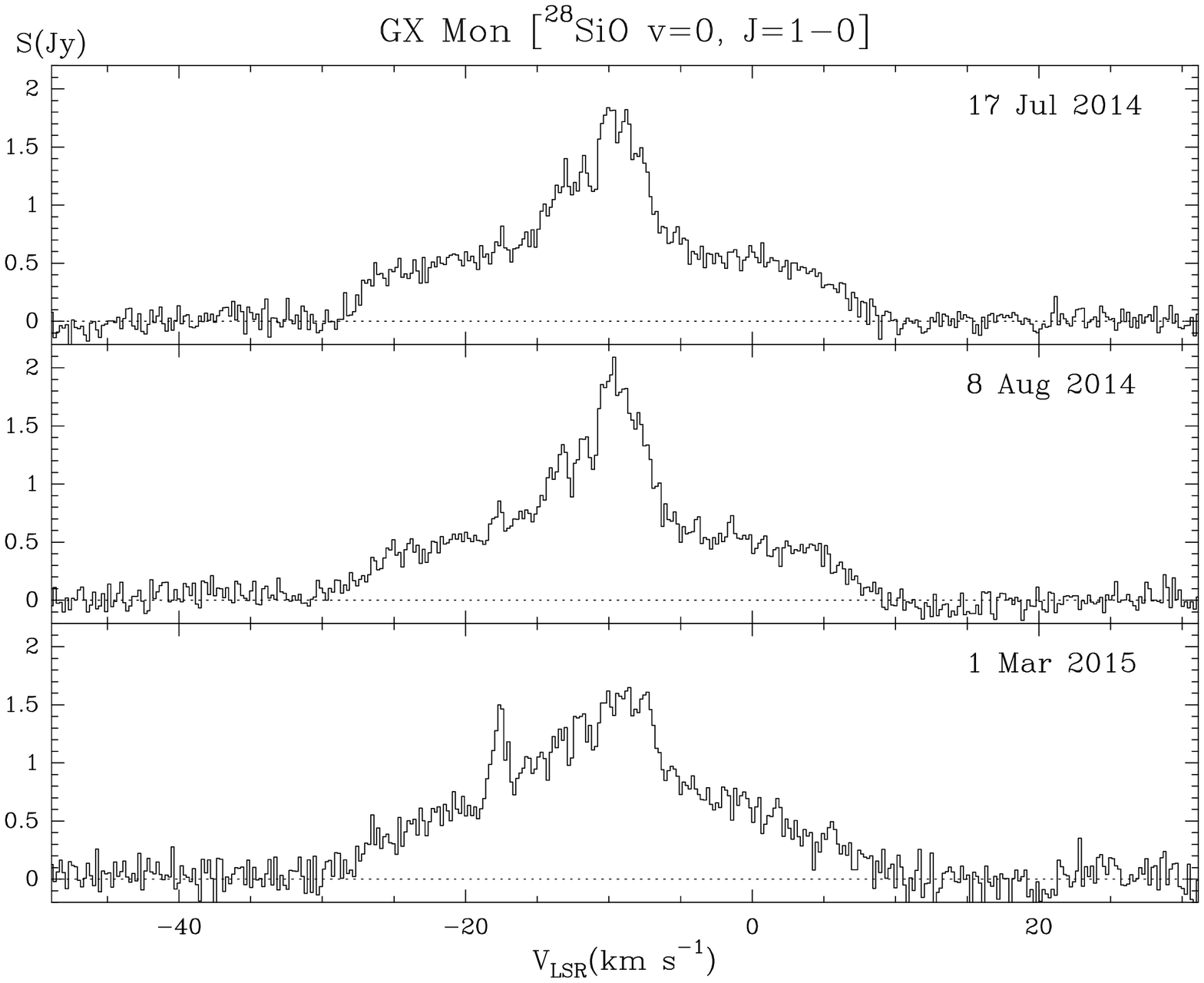}}}
   \caption{SiO $v$=0 \juc\ spectra in the O-rich Mira-type variable star
     GX Mon. The units and dates of the observations are indicated.}
    \label{f.gxmon}%
    \end{figure}

   \begin{figure}
   \centering{\resizebox{8cm}{!}{
   \includegraphics{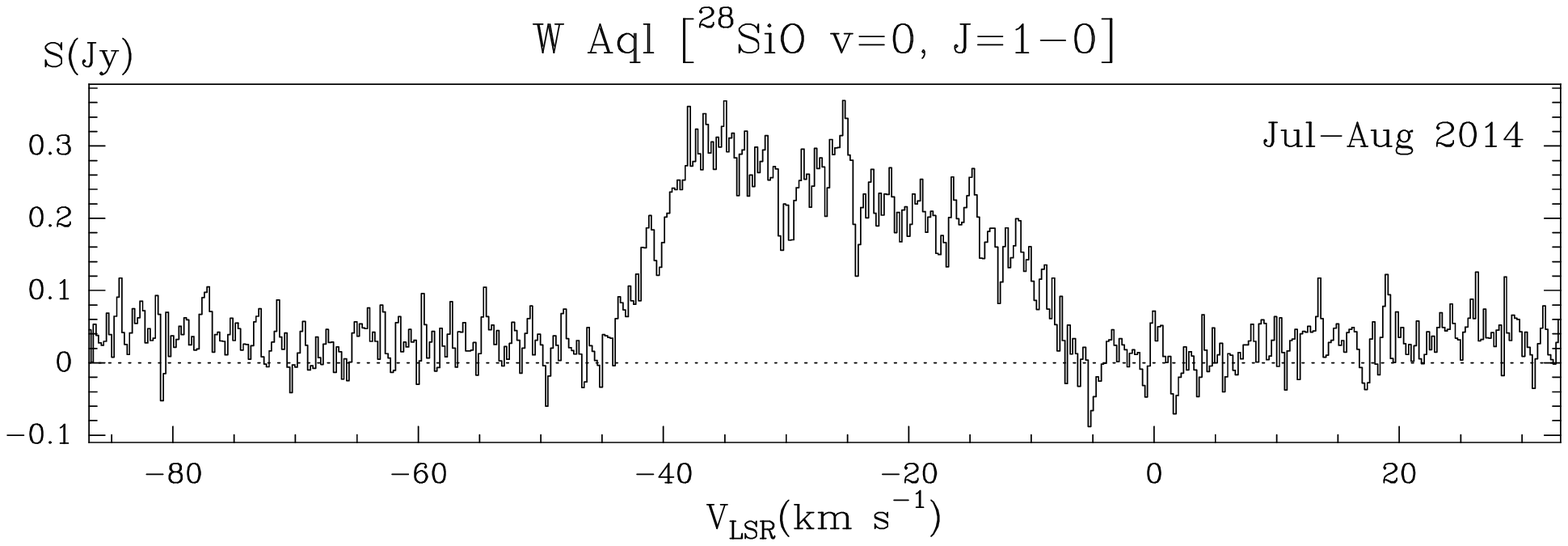}}}
   \caption{SiO $v$=0 \juc\ spectra in the S-type M-type variable star W
     Aql. The units and dates of the observations are indicated.}
    \label{f.waql}%
    \end{figure}


\begin{figure}
\centering{\resizebox{8cm}{!}{
\includegraphics{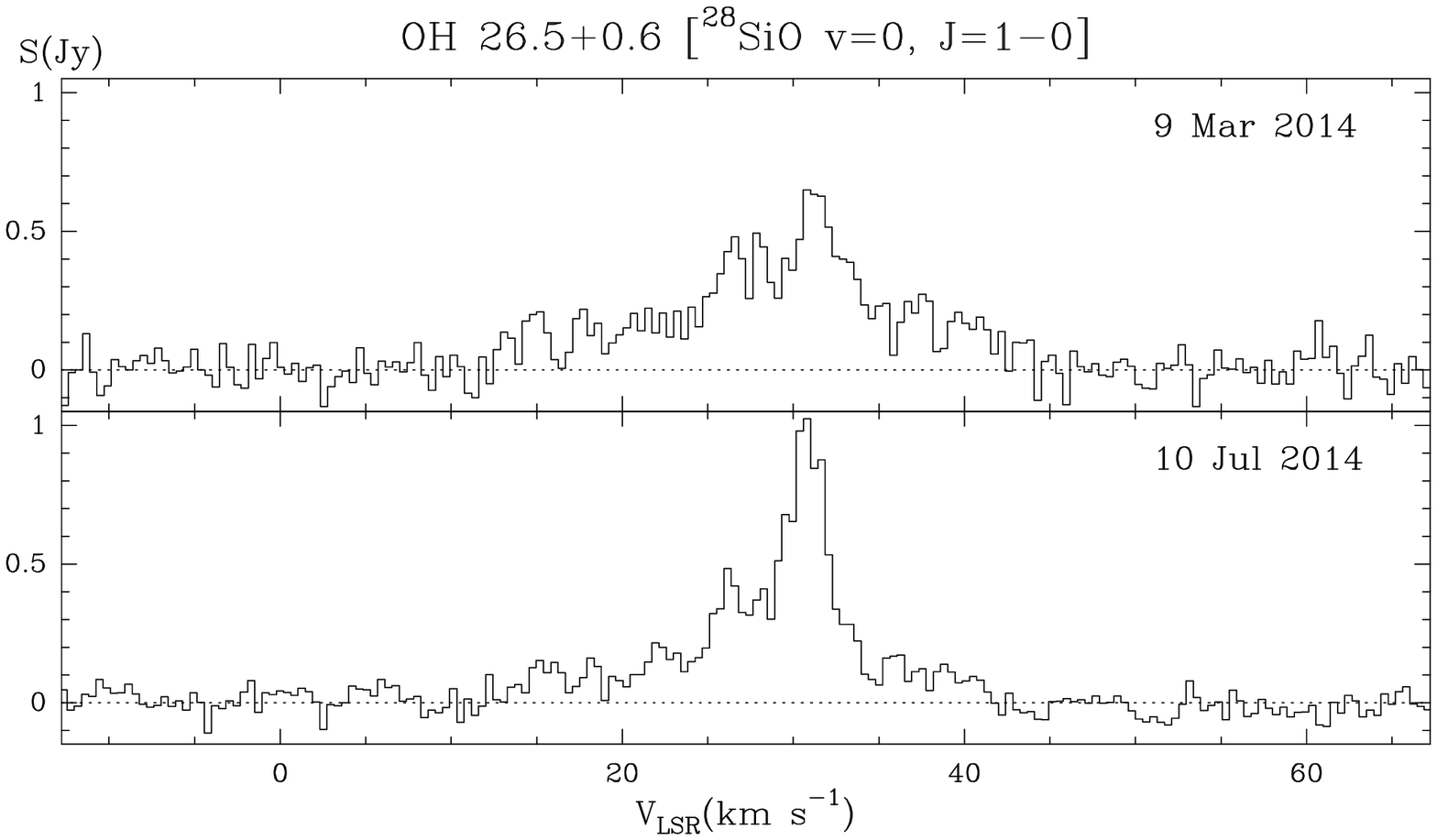}}}
\caption{SiO $v$=0 \juc\ spectra in the OH/IR star OH\,26.5+0.6. The
units and dates of the observations are indicated.}
\label{f.oh265}%
\end{figure}

\begin{figure}
\centering{\resizebox{8cm}{!}{
\includegraphics{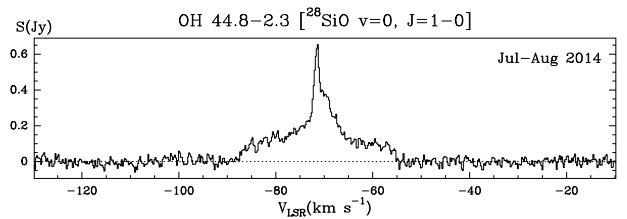}}}
\caption{SiO $v$=0 \juc\ spectra in the OH/IR star OH\,44.8-2.3. The
units and dates of the observations are indicated.}
\label{f.oh44}%
\end{figure}
\begin{figure}
\centering{\resizebox{8cm}{!}{
\includegraphics{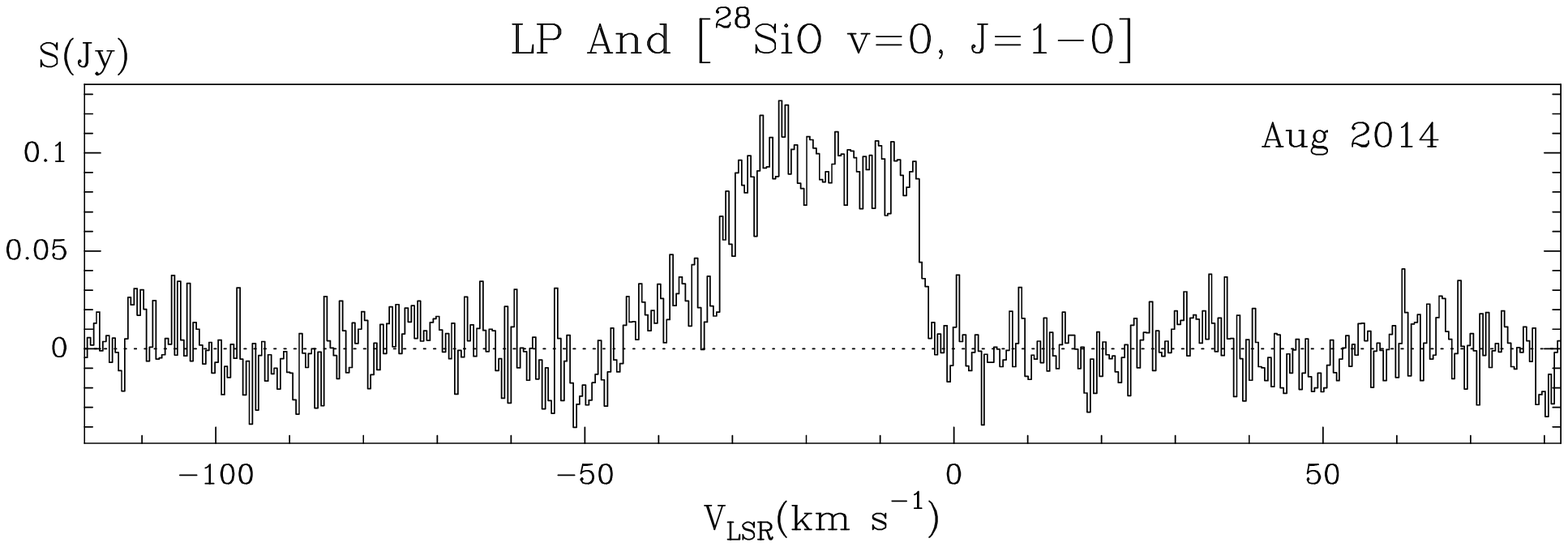}}}
\caption{SiO $v$=0 \juc\ spectra in the C-rich Mira-type variable star
LP And. The units and dates of the observations are
indicated. We note the ripples in the baseline due to instrumental
effects not corrected by our two-degree baseline removal.}
\label{f.lpand}%
\end{figure}

\begin{figure}
\centering{\resizebox{8cm}{!}{
\includegraphics{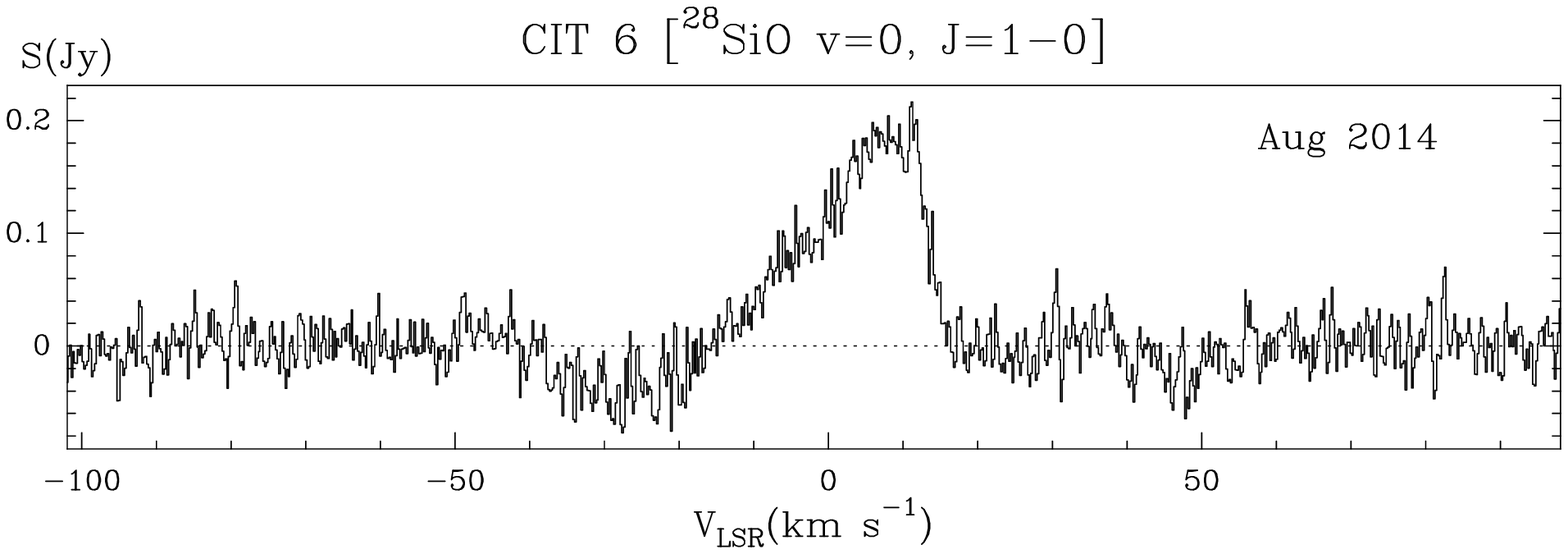}}}
\caption{SiO $v$=0 \juc\ spectra in the C-rich Mira-type variable star
CIT 6 (RW LMi). The units and dates of the observations are
indicated. We note the ripples in the baseline due to instrumental
effects not corrected by our one-degree baseline removal.}
\label{f.cit6}%
\end{figure}

\begin{figure}
\centering{\resizebox{8cm}{!}{
\includegraphics{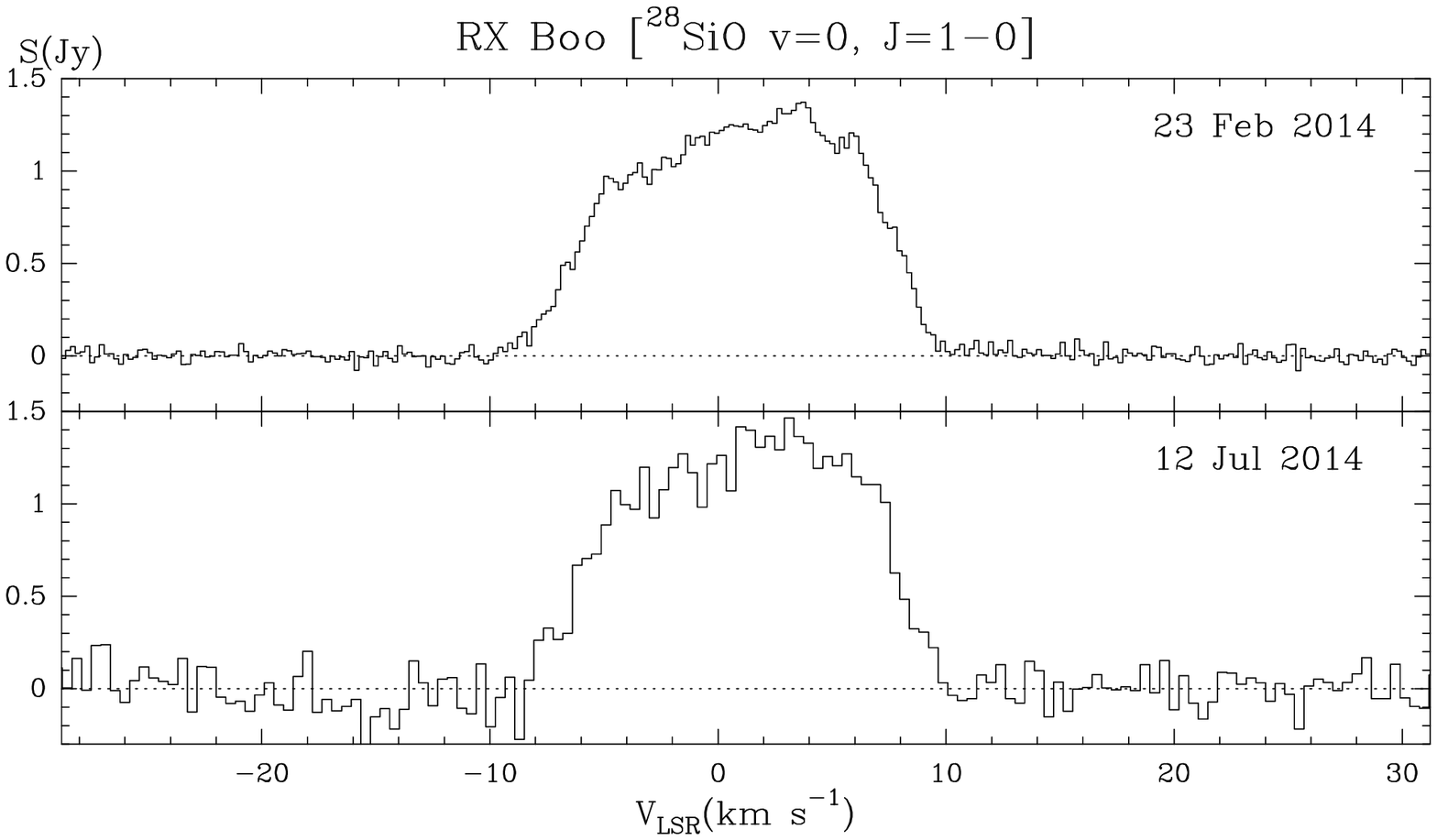}}}
\caption{SiO $v$=0 \juc\ spectra in the O-rich semiregular variable
RX Boo. The units and dates of the observations are indicated.}
\label{f.rxboo}%
\end{figure}

\begin{figure}
\centering{\resizebox{8cm}{!}{
\includegraphics{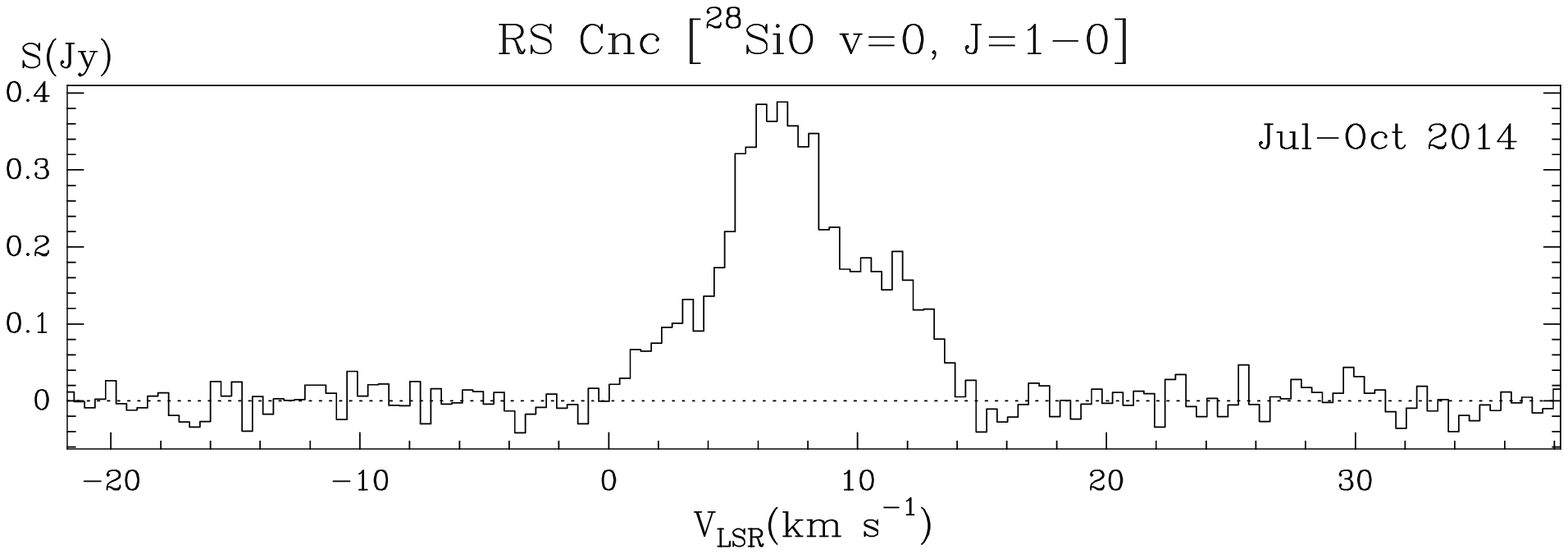}}}
\caption{SiO $v$=0 \juc\ spectra in the S-type semiregular variable star
RS Cnc. The units and dates of the observations are indicated.}
\label{f.rscnc}%
\end{figure}

\begin{figure}
\centering{\resizebox{8cm}{!}{
\includegraphics{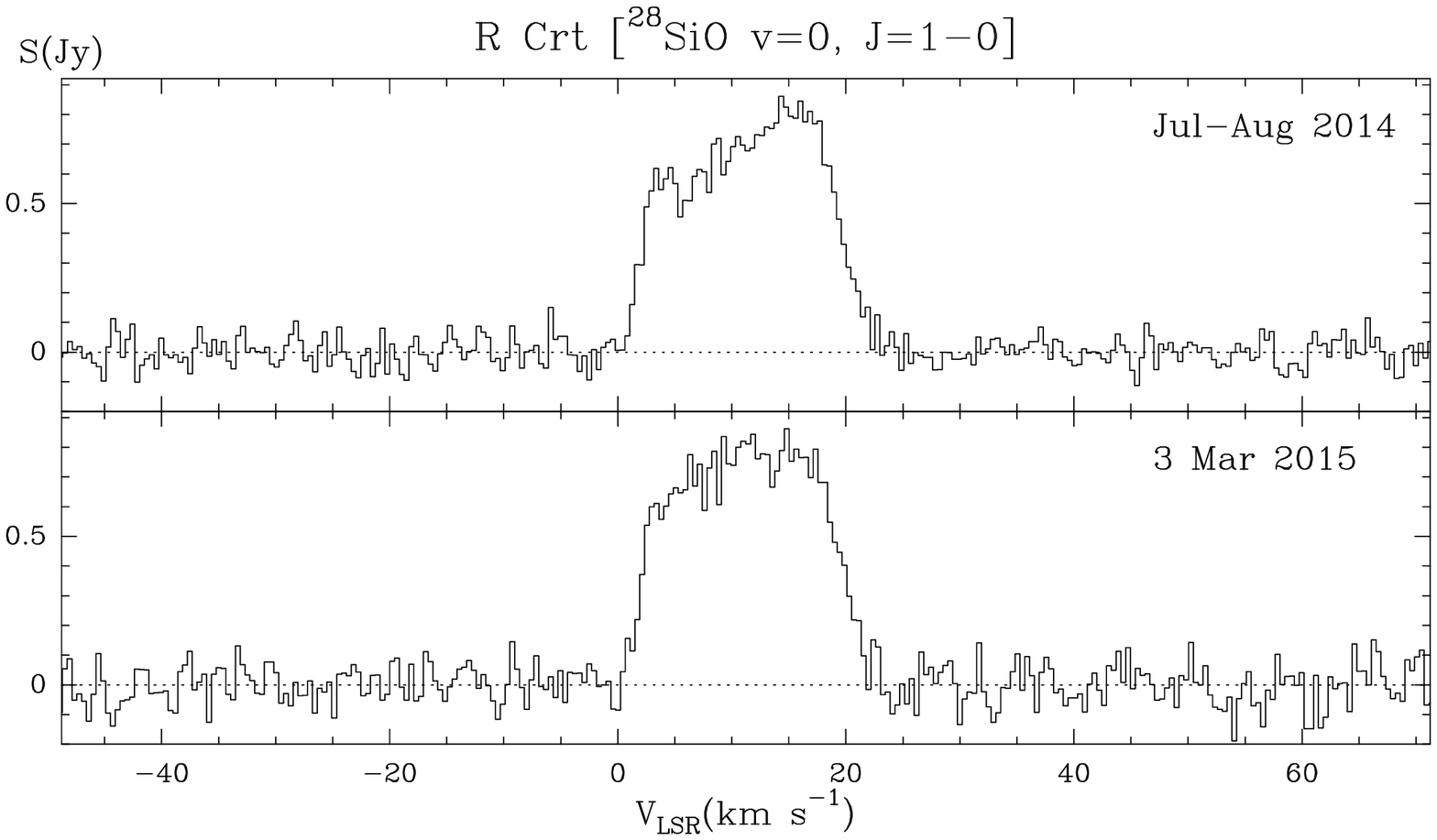}}}
\caption{SiO $v$=0 \juc\ spectra in the O-rich semiregular variable star
R Crt. The units and dates of the observations are indicated.}
\label{f.rcrt}%
\end{figure}

\newpage

\begin{figure}
\centering{\resizebox{8cm}{!}{
\includegraphics{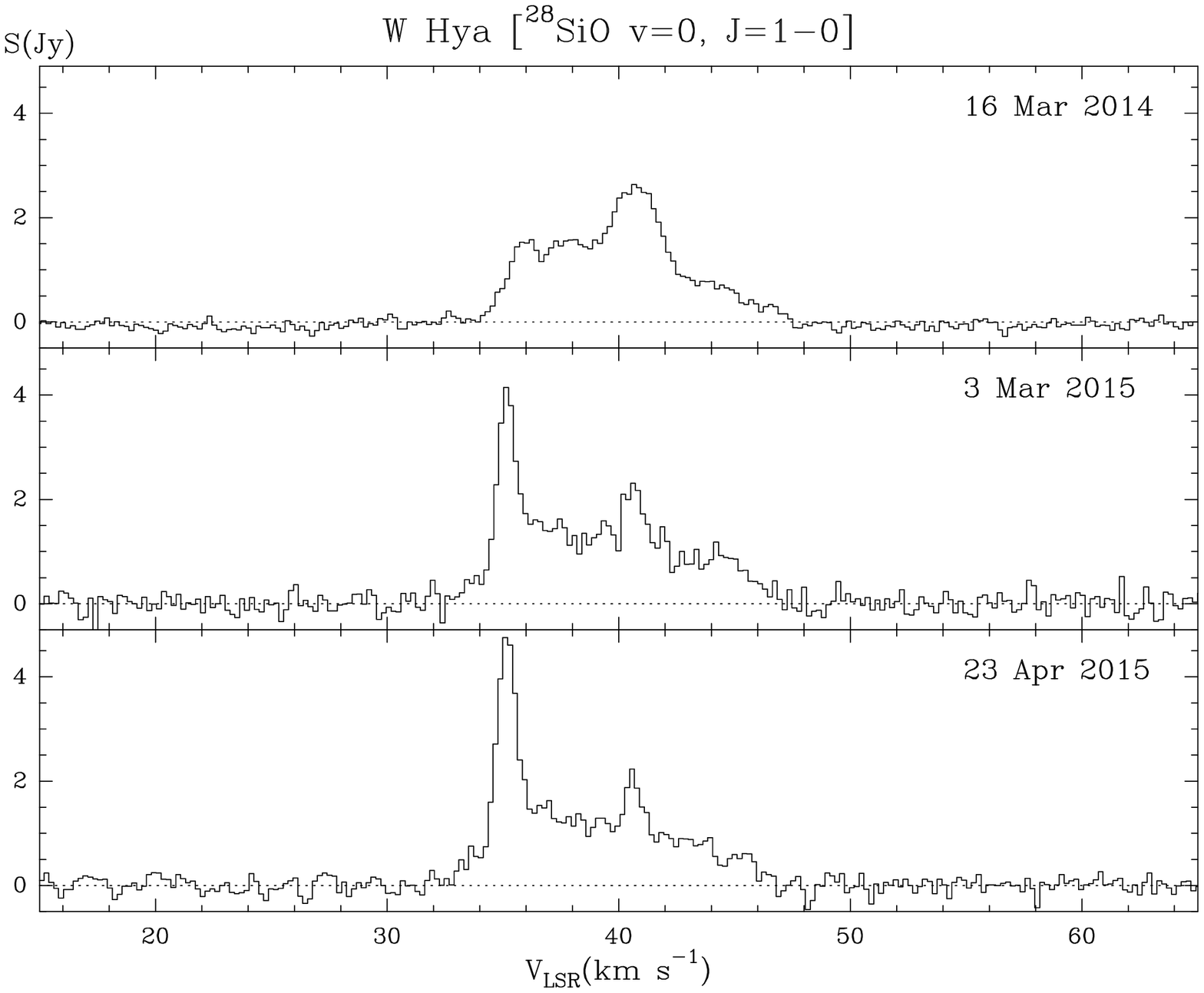}}}
\caption{SiO $v$=0 \juc\ spectra in the O-rich semiregular variable
W Hya. The units and dates of the observations are
indicated.} 
\label{f.whya}%
\end{figure}

\begin{figure}
\centering{\resizebox{8cm}{!}{
\includegraphics{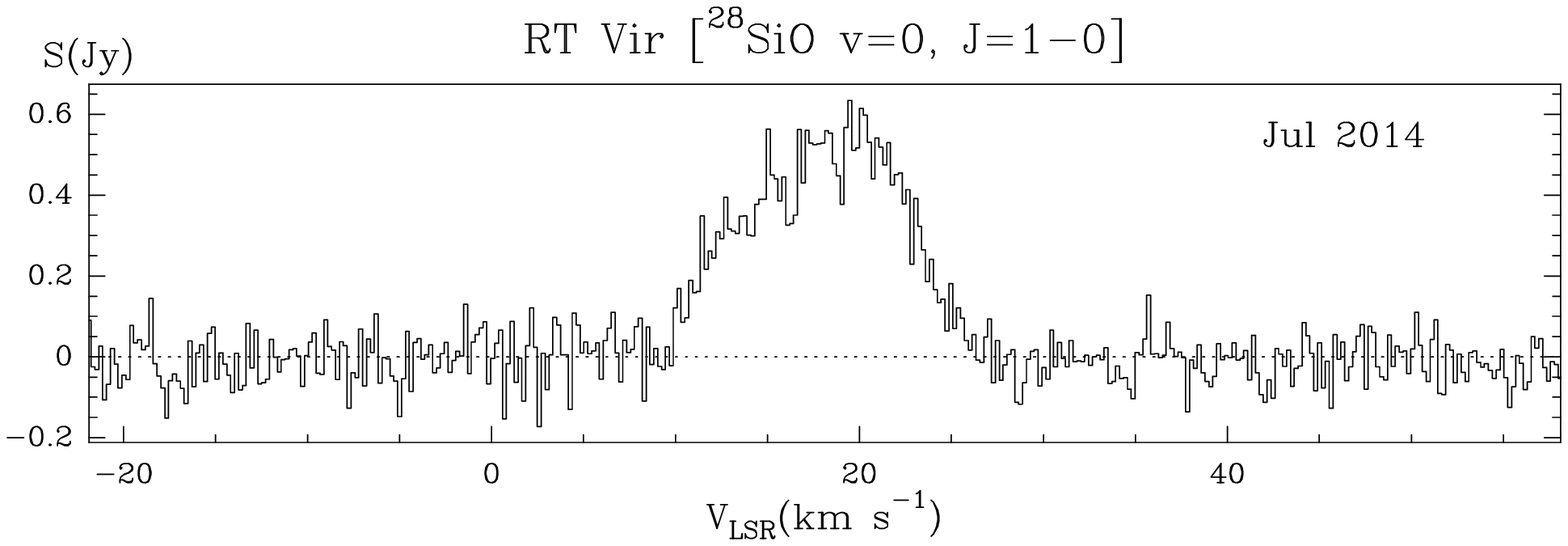}}}
\caption{SiO $v$=0 \juc\ spectra in the O-rich semiregular variable star
RT Vir. The units and dates of the observations are
indicated.}
\label{f.rtvir}%
\end{figure}

\begin{figure}
\centering{\resizebox{8cm}{!}{
\includegraphics{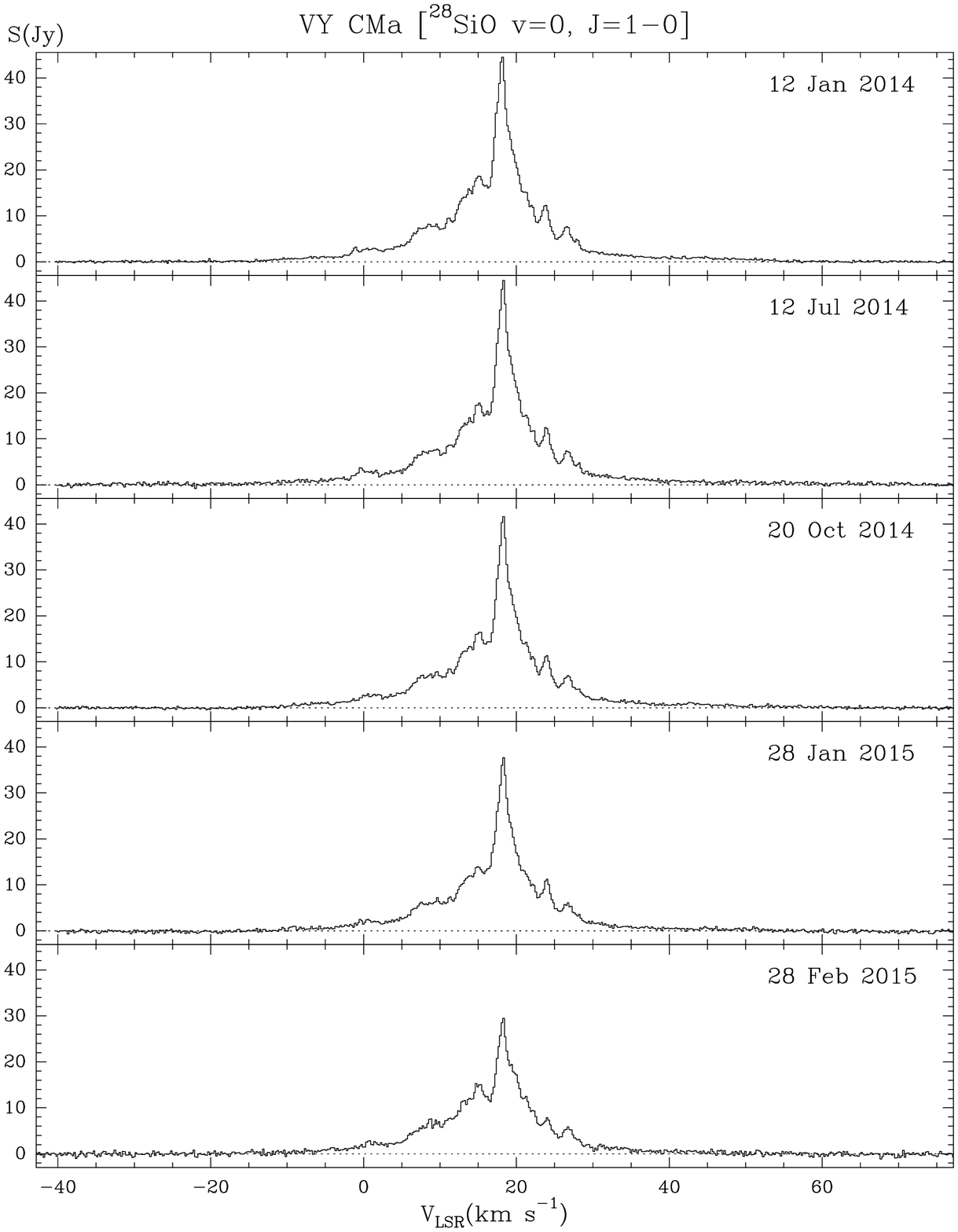}}}
\caption{SiO $v$=0 \juc\ spectra in the red supergiant star VY CMa. The units and dates of the observations are indicated.}
\label{f.vycma}%
\end{figure}

\begin{figure}
\centering{\resizebox{8cm}{!}{
\includegraphics{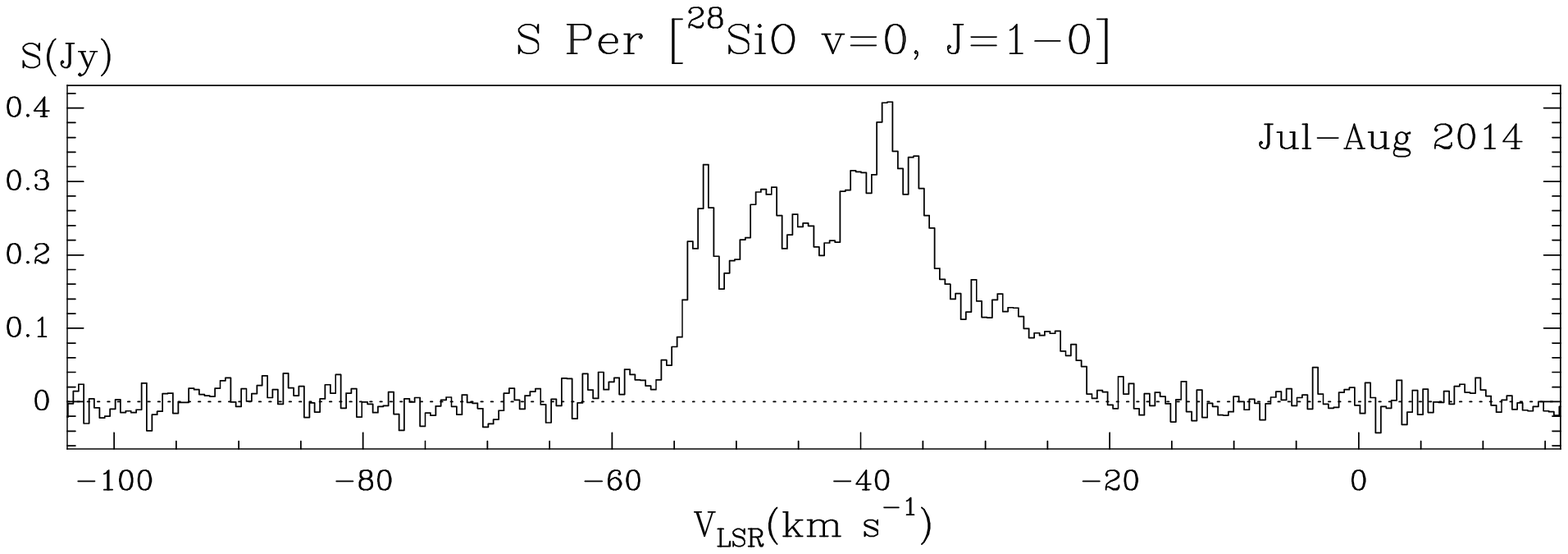}}}
\caption{SiO $v$=0 \juc\ spectra in supergiant star
S Per. The units and dates of the observations are indicated.}
\label{f.sper}%
\end{figure}
\begin{figure}
\centering{\resizebox{8cm}{!}{
\includegraphics{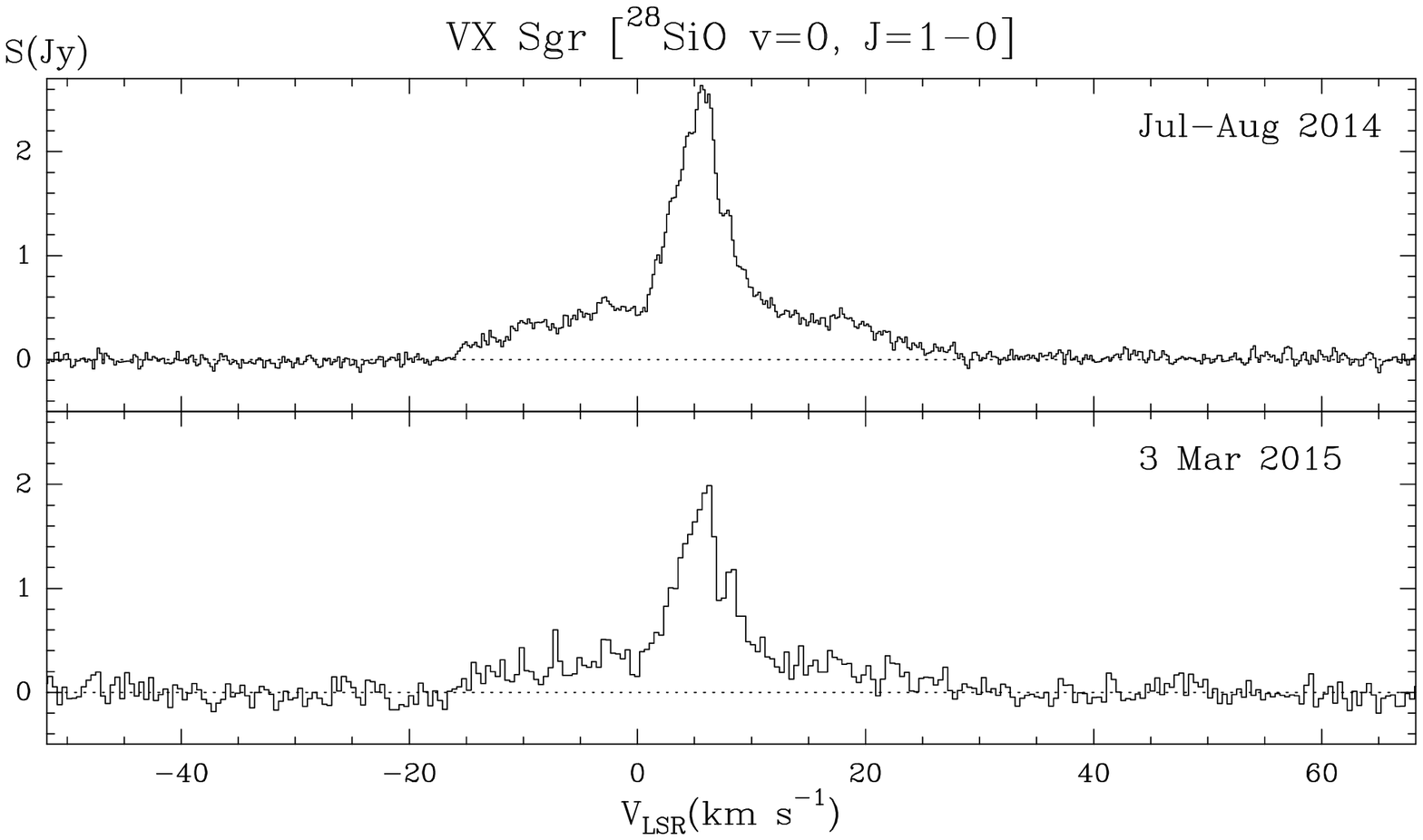}}}
\caption{SiO $v$=0 \juc\ spectra in supergiant star
VX Sgr. The units and dates of the observations are indicated.}
\label{f.vxsgr}%
\end{figure}

\newpage

\end{document}